# Roadmap on Topological Photonics

*Hannah Price[1*], Yidong Chong[2*], Alexander Khanikaev[3*], Henning Schomerus[4],Lukas J. Maczewsky[5], Mark Kremer[5], Matthias Heinrich[5], Alexander Szameit[5], Oded Zilberberg[6], Yihao Yang[7], Baile Zhang[7], Andrea Alù[8], Ronny Thomale[9], Iacopo Carusotto[10], Philippe St-Jean[11,12], Alberto Amo[13], Avik Dutt[14,15], Luqi Yuan[16], and Shanhui Fan[14], Xuefan Yin[17], Chao Peng[17], Tomoki Ozawa[18] and, Andrea Blanco-Redondo[19]*

[1] University of Birmingham, UK
[2] Nanyang Technological University, Singapore
[3] The City College of New York, USA
[4] Lancaster University, UK
[5] Institute for Physics, University of Rostock, Germany
[6] Institute for Theoretical Physics, ETH Zurich, 8093 Zürich, Switzerland
[7] Nanyang Technological University, Singapore
[8] CUNY Advanced Science Research Center, USA
[9] Julius-Maximilian University of Würzburg
[10] INO-CNR BEC Center and Dipartimento di Fisica, Università di Trento, 38123 Povo, Italy
[11] Center for Nanosciences and Nanotechnologies, University of Paris-Saclay, Palaiseau, France
[12] Université de Montréal, Canada
[13] Laboratory of Physics of Lasers, Atoms and Molecules, University of Lille, France
[14] Stanford University, USA
[15] Department of Mechanical Engineering, and Institute for Physical Science and Technology, University of Maryland, College Park, MD 20742, USA.
[16] Shanghai Jiao Tong University, China
[17] Peking University, China
[18] Advanced Institute for Materials Research, Tohoku University, Sendai 980-8577, Japan
[19] Nokia Bell Labs, USA

*Emails:

Hannah Price: H.Price.2@bham.ac.uk

Yidong Chong: Yidong@ntu.edu.sg

Alexander Khanikaev: akhanikaev@ccny.cuny.edu

## Abstract

Topological photonics seeks to control the behaviour of the light through the design of protected topological modes in photonic structures. While this approach originated from studying the behaviour of electrons in solid-state materials, it has since blossomed into a field that is at the very forefront of the search for new topological types of matter. This can have real implications for future technologies by harnessing the robustness of topological photonics for applications in photonics devices. This Roadmap surveys some of the main emerging areas of research within topological photonics, with a special attention to questions in fundamental science, which photonics is in an ideal position to address. Each section provides an overview of the current and future challenges within a part of the field, highlighting the most exciting opportunities for future research and developments.

## Introduction

*Hannah Price[1], Yidong Chong[2] and Alexander Khanikaev[3]*

[1] University of Birmingham, UK

[2] Nanyang Technological University, Singapore

[3] The City College of New York, USA

Over the last decade, topological photonics has established itself as one of the most promising platforms in which topological phases can be explored and exploited. The origins of the field reach back to the discovery of the quantum Hall effect in 1980 [1], and the consequent realization that electronic energy bands can be characterized by integer topological invariants [2,3]. Different topological phases are described by both different kinds and values of topological invariants depending on intrinsic symmetry classification of the system, and are isolated from one another by topological phase transitions. Similar to conventional phase transitions, topological phase transitions are accompanied by an abrupt change in the ground state of the system, in the latter case, however, transitions cannot be described or understood as a consequence of any spontaneous symmetry breaking. These ideas were originally developed entirely within the context of condensed matter physics, but in 2005 it was proposed that photonic crystals can exhibit a new type of electromagnetic wave, analogous to the edge states of a quantum Hall system, that travels unidirectionally and is intrinsically resistant to scattering [4,5]. Several pioneering works followed, first in microwave photonic crystals [6], and later in waveguide arrays [7], silicon photonics [8, 9], and other photonic platforms [10, 11].

The field has continued to expand rapidly, both in terms of the variety of photonic systems and the range of phenomena being explored. Excitingly, topological photonics is now firmly established at the forefront of fundamental topological physics research. The 2015 discovery of Weyl points in a 3D photonic crystal [12], for example, occurred simultaneously with the discovery of solid-state Weyl semimetals [13,14]. Other instances of topological phases that were first (or simultaneously) experimentally realized using photonics include quantum anomalous Hall insulators [6], Floquet topological insulators [7, 15, 16], higher-order topological insulators [17, 18], and topological Anderson insulators [19]. Moreover, effects that are unique to photonic systems, such as nonlinear optical effects and gain and loss in active optical media, have enabled the extension of topological physics to previously unthinkable nonlinear [20-23] and non-Hermitian domains [24-31]. Interestingly, advance of topological concepts has also enabled new understanding and interpretation of electromagnetic phenomena beyond the boundaries of topological-insulator photonic phases [32-34]. Another special property of light is that it can interact with a broad range of excitations, from excitons in semiconductors to Rydberg states in atomic gases to magnons and spin-waves in magnetic materials, opening up an even broader domain of topological polaritonics [29, 35-37]. We expect topological photonics to continue to rise in importance as interest grows in more finely-tuned topological phases of matter, aided by the fact that complex lattices with special symmetries are often easier to engineer in photonics than using real materials.

In this Roadmap, leading experts in topological photonics anticipate the most significant research directions that the field will explore during the next few years. We aim to give the reader an overview of the most promising emerging trends, as well as the principal challenges that have to be tackled. Since its inception, topological photonics has been a broad, interdisciplinary, and eclectic field, and in this spirit the choice of topics in this Roadmap is an opinionated one. The Roadmap makes no attempt to provide a comprehensive survey of the activities in the field, nor to provide a pedagogical introduction to the subject. For that, we refer the reader to the several excellent review articles in the literature [37-42].

The Roadmap is divided into several topics, grouped thematically.

The first part of the Roadmap covers the use of topological photonics to explore novel topological phases, as well as deep or exotic features of topological phases that are difficult to access in real materials. In article 1, Schomerus presents an overview of synthetic gauge fields, a concept that underlies much of the research in topological photonics, highlighting the key themes of non-equilibrium physics, nonlinearities, non-Hermiticity and quantum effects that will be further discussed in the rest of the Roadmap. The topic of non-equilibrium physics is developed in article 2 by Maczewsky, Kremer, Heinrich and Szameit, who describe the use of photonics to engineer "Floquet" band structures associated with periodic time-dependent modulations, giving rise to novel topological phases some of which have no static counterparts. Time-dependent driving is also central to the physics of topological pumping, which, as discussed by Zilberberg in article 3, provides a way for photonic experiments to explore deep features of topological boundary modes, including those occurring in unconventional lattices such as quasicrystals. In article 4, Yang and Zhang discuss developments and opportunities for 3D topological photonics, where a large number of new topological phases remain to be explored, and many interesting experimental, technological and conceptual questions remain to be addressed.

The second part of the Roadmap surveys novel phenomena that can be observed in topological photonics. Our standard theoretical understanding of band topology was developed in the context of linear non-interacting Hermitian systems, a regime that is applicable to passive photonic devices operating with classical low-intensity light. Upon dropping one or more of these conditions, the properties of photonic topological modes can be altered, sometimes dramatically and in ways that challenge long-held ideas about how topological modes should behave. In article 5, Alù discusses how topological photonics is altered by the inclusion of optical nonlinearities, a topic that poses substantial challenges and opportunities for both theory and experiment. In article 6, Thomale surveys topological photonics in the non-Hermitian regime in which electromagnetic energy is not conserved, which offers especially exciting opportunities for revising our ideas about band structure topology. Another exciting developing frontier is that of strongly-interacting topological photonics, presented by Carusotto in article 7, in which the combination of single-photon nonlinearities and synthetic magnetic fields can lead to exotic types of strongly-correlated many-body phases, such as fractional quantum Hall states of light. This part of the Roadmap concludes with articles on two emerging topological photonics platforms that are especially amenable to realizing novel phenomena. In article 8, St-Jean and Amo discuss topological microcavity polaritons, which exploits the tunability and interactions offered by these hybrid light-matter quasiparticles. In article 9, Dutt, Yuan and Fan introduce synthetic dimensions for photonics, in which additional degrees of freedom are harnessed and coupled together so as to artificially simulate the effects of extra synthetic spatial dimensions, allowing for the engineering of novel topological models.

The third and final part of the Roadmap covers the possible applications of topological photonics. The robustness of topological modes against various forms of disorder, such as fabrication imperfections, is an attractive feature for those seeking to develop practical photonic devices; since the earliest days of the field, it has been envisioned that topological modes could eventually be used in on-chip optical isolators or optical delay lines in integrated photonics devices. Taking a broader view, the physics of topological phases has, despite its elegance and multiple decades of research, found scant applications to date beyond metrology [43]; photonics may very well be the setting in which these concepts are first turned into practical technologies. In this Roadmap, we have selected three long-term research areas that hold great promise for future applications. In article 10, Yin and Peng discuss topological bound states in the continuum, an emerging approach to light-trapping with intriguing applications in lasers, filters and sensors. In article 11, Carusotto and Ozawa describe topological lasers, in which the use of topological boundary modes for lasing leads to properties that may be superior to those of conventional lasers. Finally, in article 12, Blanco-Redondo explores how topology can be used to protect quantum states and entanglement, and possible applications in quantum computing and quantum information with light.

Topological photonics is a flourishing field and a leading platform for the exploration of new types of topological effects, which are hard to realise in solid-state materials. Beyond condensed matter, there is also a synergy between topological photonics and other fields, including cold atoms and classical systems, such as acoustic and mechanical metamaterials, active fluids and electrical circuits. Each of these falls into a different niche in terms of how the topological properties can emerge and what features can easily be integrated into the system. For example, acoustic metamaterials are especially easy to fabricate, making this platform well-suited to proto-typing, while the essentially arbitrary connectivity possible in electrical circuits is ideal for simulating highly inter-connected topological lattices. Amongst these, topological photonics has the advantage of combining different strengths, including being highly tuneable and easy to implement in experiments, as well as giving access to both the classical and quantum domain. Moreover, topological photonics is almost certainly the most likely to move beyond “proof-of-principle” experiments in the future, towards real applications. The potential broad impact of topological photonics stems largely from the new mechanism of confinement and control of electromagnetic field in the form of boundary states and from their intrinsically robust nature, which envisions a wide range of resilient devices operating from microwave to optical frequencies in both passive and active regimes. With this potential, the future for topological photonics looks bright, with exciting opportunities and challenges ahead both for fundamental science and applications.

**Acknowledgements**

HMP is supported by the Royal Society via grants UF160112, RGF\EA\180121 and RGF\R1\180071. CYD acknowledges support from the Singapore MOE Academic Research Fund Tier 3 Grant MOE2016-T3-1-006, and the National Research Foundation Competitive Research Programs NRF-CRP23-2019-0005 and NRF-CRP23-2019-0007. ABK acknowledges support by the ONR award N00014-21-1-2092, the NSF grant DMR-1809915, and the Simons Collaboration on Extreme Wave Phenomena.

## 01- Synthetic gauge fields

*Henning Schomerus*

Lancaster University, UK

### Status

Much of traditional optics relies on the refractive index to guide and manipulate light. This provides a mechanism to modify the optical path length, resulting in systematic interference effects based on phase differences, as well as a mechanism to attenuate or amplify light, resulting in effects based on the light intensity. However, with the refractive index being a scalar field, both mechanisms are constrained by optical reciprocity, which inhibits media to exhibit a directed response. In particular, many of the most coveted topological effects–such as embodied, e.g., in the Quantum Hall effect (QHE)–rely on nonreciprocal transport, and therefore require vector potentials [1]. On the fundamental level such potentials are realised by dynamical gauge fields that couple to charged matter; the EM field is one such example based on the U(1) degree of freedom [2]. Synthetic gauge fields are effective means that induce a nonreciprocal response into light itself. Being freely engineered, they are also involved in defining the effective dimensionality and symmetries of a system, provide the topologically distinct configurations as well as the defects and interfaces that pin topological modes, and can act on internal degrees of freedom provided by polarisation or discrete modes–features that all are instrumental in determining the universality class of a systems [3]. As such, these fields provide the essential background on which topological physics plays out.

Synthetic gauge fields were pioneered in atom-optical settings, where rotations induce a Coriolis force and spatially patterned atom-laser interactions induce phase shifts that replicate the effect of a magnetic flux [4,5]. Photonic platforms offer a wide range of solutions (see Fig. 1). Following the early proposal by Haldane and Raghu [6], the conceptually most direct implementations rely on magneto-optical effects utilizing gyrotropic materials [7], or exploit the polarization degree in media that exhibit circular birefringence [8] or bianisotropy [9]. A technologically robust engineered solution is the implementation of coupled ring resonators in which the opposite propagation components are well separated [10], resulting in two internally nonreciprocal sectors that are related by time-reversal symmetry and in which the dynamical phase picked up along resonator segments acts as a pseudo-gauge field of opposite signs. Pseudomagnetic fields can also be induced by distortions of photonic lattices that possess Dirac cones at finite wave vectors [11,12,13]. Reinspecting how gauge fields enter effective Hamiltonians with regards to four-dimensional space-time, nonreciprocity can furthermore be induced by temporal variations [14,15,16]. In wave-guide arrays, time can be replaced by the propagation distance, with fields again displaying opposite signs for the two propagation directions [17]. With the help of intrinsic degrees of freedom, it is possible to attain more complex scenarios involving nonabelian fields [18]. By elevating such freedoms to synthetic dimensions and combining them with temporal variation, it is possible to explore higher-dimensional analogues of topological physics, such as the 4-dimensional QHE [19]. Interaction effects are incorporated in exciton polaritons [20,21] and circuit QED [22], where bosonic excitations are manipulated through their coupling to electrons exposed to magnetic fields or Josephson-junction rings penetrated by magnetic fluxes. Nontrivial synthetic background potentials can also be inherited from electronic topological materials, for instance, when their optical response reveals an emergent axionic field [23].

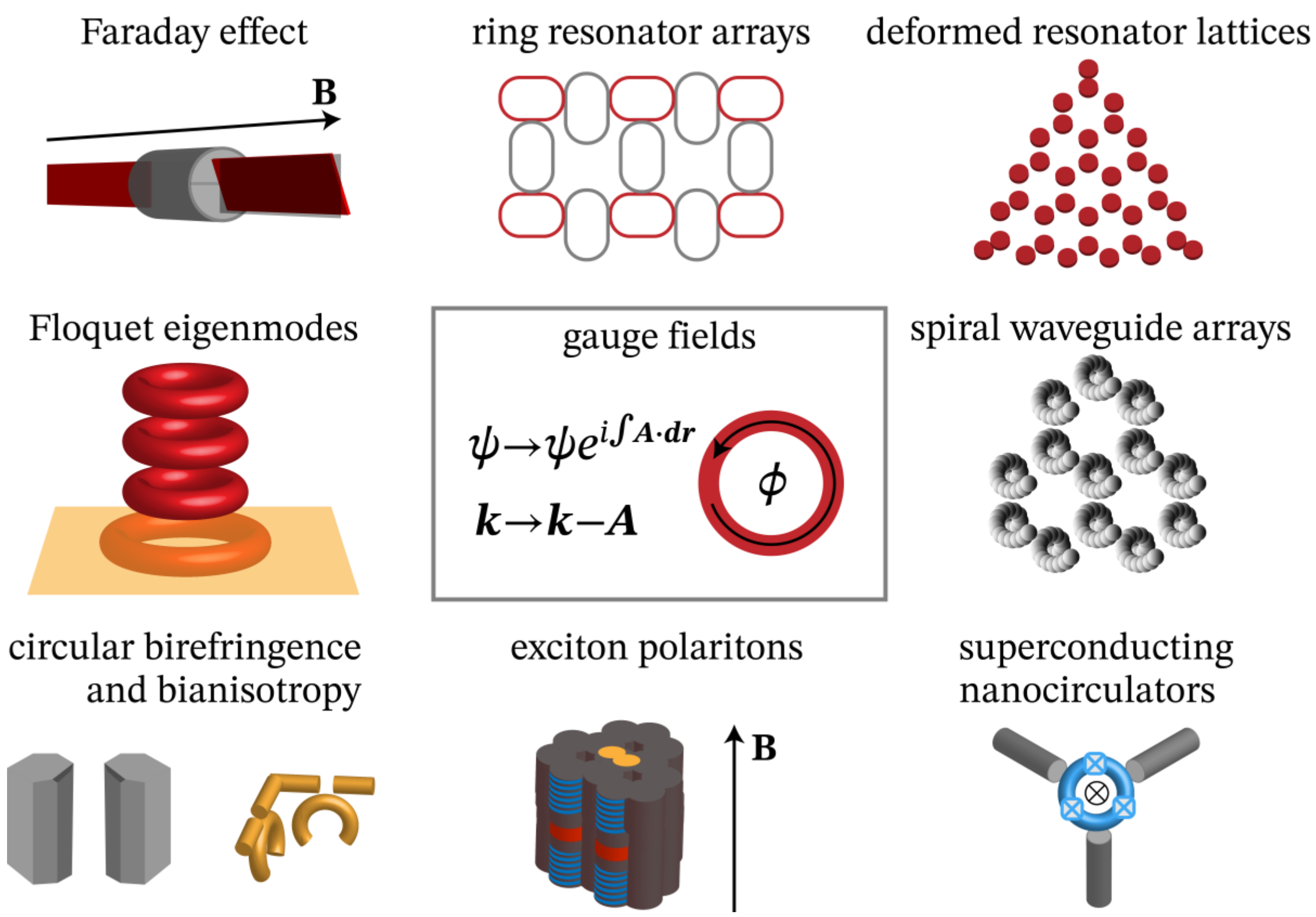

**Figure 1.** Gauge fields induce nontrivial phase shifts for round trips along closed loops, which in effective models correspond to a shift of the wave vector. Synthetic gauge fields replicate these shifts by a variety of means, including magneto-optical effects such as the Faraday effect in gyrotropic materials, path-length differences in ring resonator arrays with suppressed backscattering, strained Dirac materials with suppressed intervalley scattering, temporal variations in Floquet systems or quantum walks, analogous variations along the propagation direction of waveguide arrays, birefringence and bianisotropy, and interactions with other degrees of freedom, as in exciton polaritons or nano-circulators incorporating superconducting Josephson junctions.

## Current and Future Challenges

Topological phenomena promise robustness independent of details, but also rely on strict adherence to overall constraints that place a system into a desired universality class. All mentioned platforms offer unique opportunities, where discrete intrinsic degrees of freedom as well as the dissipative and nonequilibrium physics and nonlinearities attained through material gain and loss or steady-state pumping alter the mathematical base space, and hence the fundamental nature of topological phenomena. The conceptual understanding of the interplay of these effects is still developing, both in terms of topological effects that survive as well as novel effects that emerge, which defines a key challenge to unlocking the complete potential of topologically robust photonic phenomena [24].

Precisely because of this link, all these effects need to be carefully controlled, and undesirable features suppressed. In terms of implementations, progress is arguably most advanced for the case of synthetic fields based on decoupled propagation sectors [10], where backscattering needs to be negligible, or deformed photonic lattice and resonator systems [11,17,13], which rely on optical elements with identical characteristics that furthermore may need to be positioned precisely, or coupled selectively if a sublattice symmetry is to be obeyed. With present technology and materials, intrinsic magneto-optical effects [6,7,8,9] can be appreciable at microwave frequencies, whilst their magnitude at optical frequencies is often limited. Considerable technological challenges persist in settings that involve active and dissipative media [20,21] and work towards quantum-limited operation [22], where lifetime broadening and noise constitute particular roadblocks to realize genuinely robust topological bandstructure phenomena. For instance, lifetime differences within a topologically induced dispersion branch [25] directly impact on directed transport in a much more profound way than encountered, e.g., for spectrally isolated zero modes [26]. On the other hand, combining synthetic fields with suitably distributed gain and loss can result in directed couplings [27], which is a key prerequisite for directed amplification.

While static solutions are desirable and sufficient for many applications, there is significant merit in elevating synthetic gauge fields to independent dynamical entities. This long-term challenge emerges when one steps back and revisits their field-theoretical foundations [2]. Particles exhibit topological phenomena whenever they move in nontrivial background potentials, which determine the distinct configurations as well as the defects and interfaces that pin the topological excitations. In their original field-theoretic inception, the high-energy settings of the 1970's, these background potentials were provided by fundamental gauge fields exhibiting their own dynamics, with the coupling to the particles dictated by gauge symmetry. The coupling of these fields to the matter translates the theoretical topological anomalies into a unique physical response, such as directed currents, and the dynamics of the fields elevates topological excitations into genuinely particle-like entities, that then inherit unconventional properties such as fractionalised charges or spin-charge separation. Replicating these aspects for bosonic degrees of freedom promises to advance the fundamental understanding of these systems and offers a path to presently completely unexplored, possibly highly unconventional applications.

**Advances in Science and Technology to Meet Challenges**

The scope of topological effects induced by synthetic fields can be significantly advanced by a systematic extension and translation of present mathematical classifications–still mostly routed in spectral considerations of linear, noninteracting settings but already incorporating gain and loss [28]–into a comprehensive, genuinely physical framework phrased in terms of response and transport characteristics, in analogy to the achievements in terms of conductance properties in electronic insulators and superconductors [1,3]. This framework will serve to determine which topological phenomena survive and emerge in the combination of different effects (see Fig. 2), and how they can be detected. To be suitable for photonic applications, besides gain and loss the framework will have to take into account the possibilities of nonlinearities and nonequilibrium physics, which can induce dynamical symmetry breaking [29] and instability [30]. Of practical importance is also the understanding of material dispersion and possibly nonlocal response functions, while on the quantum level relevant aspects include quantum noise [31] as well as fundamental constraints such as arising from causality [32]. Advancing this framework will provide a reliable basis for the development of device concepts with clearly defined physical requirements. Some of these requirements are bound to remain technologically challenging, with improvements in the design and control of systems required in particular for the attainment of quantum-limited operation and strong-coupling regimes. The challenges in the design of such devices could also be addressed by utilizing the gauge freedom inherent in the definition of the fields. However, this freedom can be broken by the interplay with other effects, an understanding of which therefore needs to be developed.

Fundamental advances are also critical to elevate synthetic gauge fields to independent dynamical entities. An understanding is required of how the topological physics changes where these fields couple to bosonic degrees of freedom, as this removes both the Pauli blocking from the Fermi sea as well as particle-number conservation. On the other hand, these systems can exhibit collective bosonic physics such as condensation, can again be enriched by gain, loss, and nonlinearities, and serve as a platform for topological physics in strongly coupled and interacting setting.

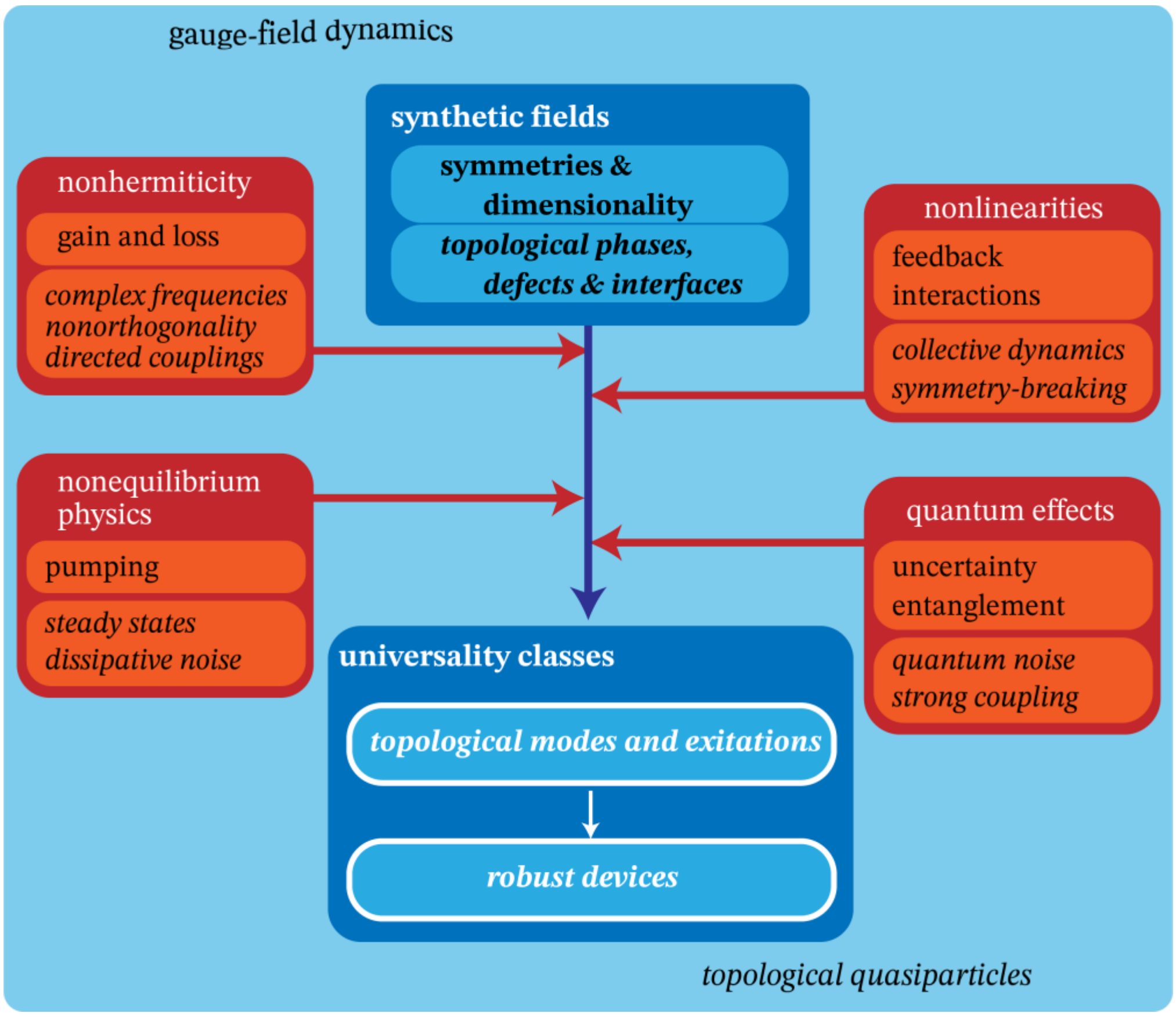

Figure 2. Synthetic gauge fields provide the background on which topological photonic physics plays out. They interact with a wide range of effects that determine the nature of topological modes and excitations. A common characteristic of engineered solutions is the lack of independent gauge-field dynamics.

## Concluding Remarks

Gauge fields provide the indispensable background on which topological physics plays out. In photonic systems, this physics is liberated from fundamental constraints and enhanced by a wide variety of effects ranging from gain and loss over nonlinearities to nonequilibrium phenomena, and occurring on the classical and quantum level. This richness offers substantial unexplored opportunities, where conventional topological effects can be enhanced and new effects emerge, but also poses significant technological challenges, when some of this physics drives the system out of the desired universality class. Overcoming these challenges will help to solidify the foundations of topological photonics as a whole, resulting in both a deeper conceptual understanding as well as in well-designed implementations of robust functional devices.

**Acknowledgements**

The author acknowledges funding by EPSRC via grant Nos. EP/J019585/1 and EP/P010180/1, and Programme Grant No. EP/N031776/1.

## 02 – Floquet topological photonics

*Lukas J. Maczewsky, Mark Kremer, Matthias Heinrich, and Alexander Szameit*

Institute for Physics, University of Rostock, Germany

### Status

The first demonstration of a Floquet topological insulator (FTI) [1] in 2013 marked a milestone in the study of topological photonics. In this seminal work, Duncan Haldane's idea of transcending the need for magnetic fields by breaking time reversal symmetry [2] was implemented with helically shaped waveguides in a femtosecond laser written photonic lattice. These experiments unequivocally confirmed that the periodic modulation indeed is capable of inducing a bandgap that hosts unidirectional edge states, a hallmark feature of topological insulators.

Importantly, the concept of periodic modulation is at the core of every FTI, where it serves as a convenient tool to tailor or entirely break time-reversal symmetry. Based on the periodicity in energy provided by the Floquet theorem, an additional degree of freedom is established in FTIs, which can yield additional protected edge states and host of other intriguing phenomena. Building upon the initial idea, a variety of novel and previously inaccessible topological phenomena came into the reach of photonics. Based on an arrangement of helically shaped waveguides with two different phases it was e.g. possible to observe Weyl points in the band structure [3]. Floquet topology enabled the emulation of two-component relativistic fermions in an optical Floquet lattice, which allowed for the study of fundamental research topics.

Moreover, FTIs were utilized to explore Fermi arcs and PT-symmetric lattices in the context of topological photonics [4]. Another prominent type of FTIs, so-called anomalous FTIs, support topological edge states despite a vanishing Chern number [5,6]. In addition, Anderson TIs, where the topological phase is brought about by the disorder, were realized [7].

In direct analogy to solid state physics, topological effects in photonics usually express themselves through unidirectional transport in real space. However, in a remarkable work, a synthetic FTI was realized with unidirectional edge states placed in reciprocal space [8]. In this vein, the salient feature of topological protection was transferred to the spectral domain, where it may enable various technological applications.

Currently, FTIs are predominantly accessible on three major platforms: femtosecond laser written waveguide arrays [9], silicon photonics [10] and ultra-cold trapped particles [11]. Therefore, the protected transport mechanism of these systems allows the support of a wide range of physical experiments.

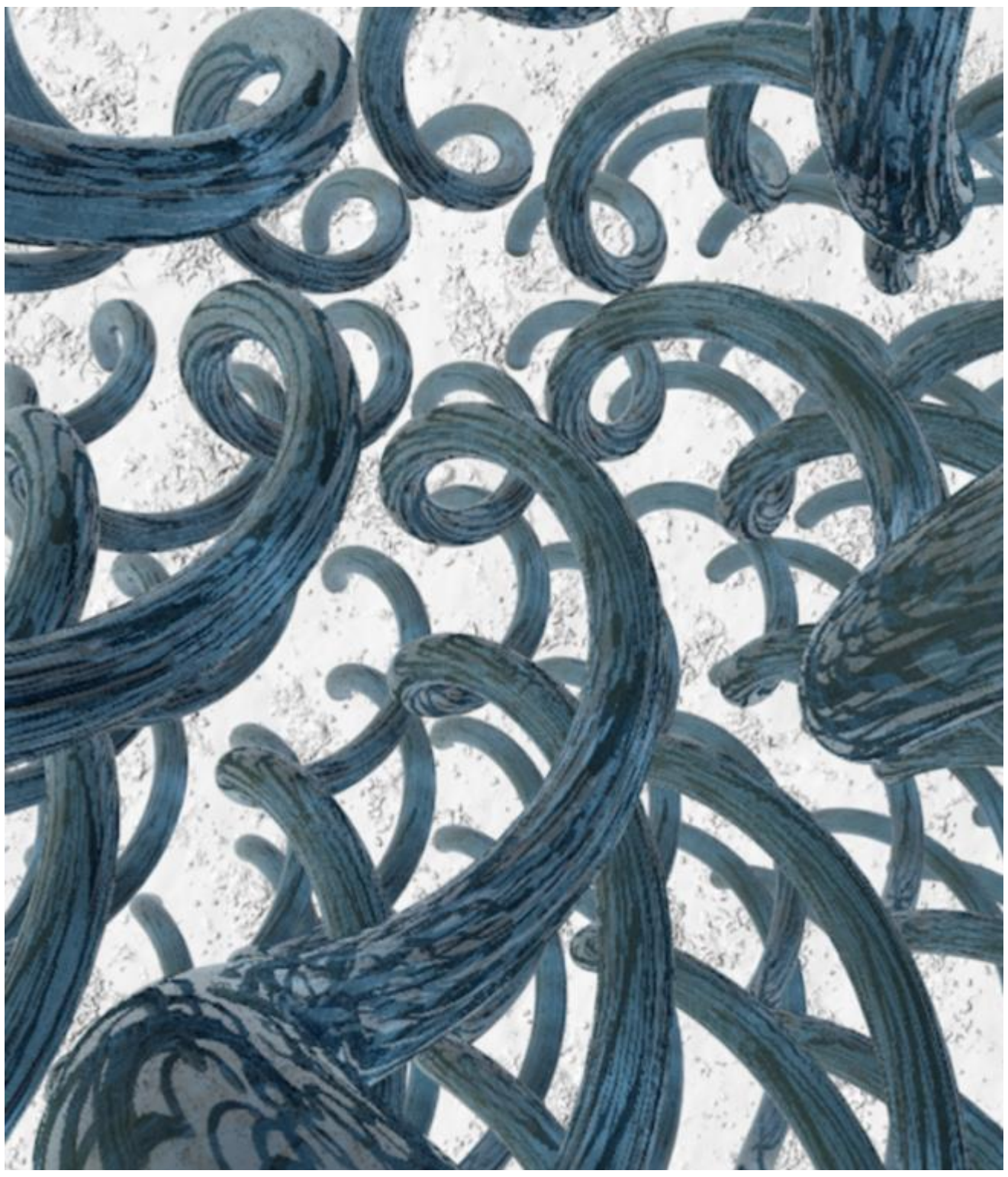

Figure 1 **Helically shaped waveguides.** The broken time reversal symmetry by virtue of helicity was the key feature of the first FTI.

## Current and Future Challenges

Although the past decade has witnessed Floquet topological photonics becoming one of the most dynamic fields of research, several issues are still under investigation. Mainly, the influence of propagation losses constrains the design and potential applications of FTIs. While the experimental observations indeed confirm the overall theoretical predictions, a derivation of a topological theory that rigorously includes the propagation losses remains an ongoing challenge. Especially for experiments that combine topology with the quantum nature of light, this will be a decisive factor. Several experimental approaches study quantum optical phenomena in non-trivial topological structures in order to utilize the topological protection for entangled states [12,13]. However, a non-trivial topological phase solely originating from quantum physics has yet to be found. The intensively studied entangled photon experiments, combined with a photonic FTI on waveguide basis, could provide further insights on how topology impacts the dynamics of quantum systems.

Furthermore, the investigation of non-Hermitian Floquet topological systems is currently in the scope of researchers. The deliberate introduction of loss mechanisms enables the study of entirely passive PT symmetric systems. In this context, the dynamics around the exceptional points are of particular interest. By realizing a Weyl exceptional ring, researchers were able to observe topological transitions in waveguide lattices [14]. However, exploring the impact of non-Hermiticity on the topological properties that go beyond the known Hermitian classifications are an ongoing challenge.

A similar problem is posed by investigating the interplay of nonlinearity and topology, since conventional topological models are fundamentally based on Bloch's theorem, i.e. a purely linear concept. Since a general definition of a band structure does not exist in the nonlinear regime, the question of how to define a suitable topological invariant has proven a tough nut to crack.

Nevertheless, the absence of a rigorous theory has not prevented initial experimental investigations that harness FTIs to demonstrate topologically protected transport in a nonlinear regime [15].

The perhaps most important step in bringing photonic FTIs to industrial applications is miniaturization and standardization across different platforms and investigating the subwavelength dynamics. By utilizing minimized structures, the interplay of topology and graph theory can be used to realize neuromorphic networks [16,17]. In particular, FTI structures could be combined with machine learning approaches and implement quantum computers as well as quantum simulation devices [18,19].

**Advances in Science and Technology to Meet Challenges**

In the next decade FTIs are facing the challenge to implement ever-more complex waveguide trajectories in order to observe distinct topological phases such as a photonic Z2 topological phase [20]. Furthermore, for industrial applications of FTIs, a suitable, stable and readily-adaptable platform will be essential. Along these lines, polymer waveguides obtained by a 3D printing technique could be a promising candidate [21]. Due to the increasing prevalence and standardization of 3D printing processes, this method can be soon expected to routinely achieve extreme precision and process stability. By combing 3D printed scaffolds with suitable filling liquids, the physical size of the individual building blocks of FTIs can be greatly reduced while maintaining the required delicate control over the couplings. As an added benefit, this technique readily allows for the introduction of higher spatial waveguide modes as potential path towards higher-dimensional FTIs.

Another promising candidate is the silicon platform, which possesses the potential for an extension of (2+1) dimensional FTIs to (3+1) dimensions, which could allow the investigation and application of one- and two-dimensional topological protected edge modes. Furthermore, the platform could host topological insulator lasers [22], which utilize the benefits of FTIs to enrich the portfolio of accessible and individual laser sources.

In order to achieve an extreme stability of FTIs, fibre loop setups [23] could serve as a promising candidate. Since it is based on the highly developed standards of the fibre-telecommunication industry it is easily adaptable and overcomes the issue of propagation losses by using advanced fibres, as well as established fibre-amplifiers.

Current FTI driving protocols are designed to verify the topological properties, however for specific applications, e.g. in quantum computing, different, and, above all, individually optimized sequences will be needed. Machine learning and neural network based processing offer powerful tools for the design of the next generation of FTIs.

Among all these potential industrial applications, FTIs will continue to serve as versatile testbed for fundamental research of the physical community, ranging from time crystals and FTIs in chalcogenide glasses to investigating many-body problems of condensed matter by synthetically implementing their fermionic properties. The additional degree of freedom in energy/time render FTIs one of the most versatile types of TIs and holds the promise for a broad range of scientific and technological breakthroughs.

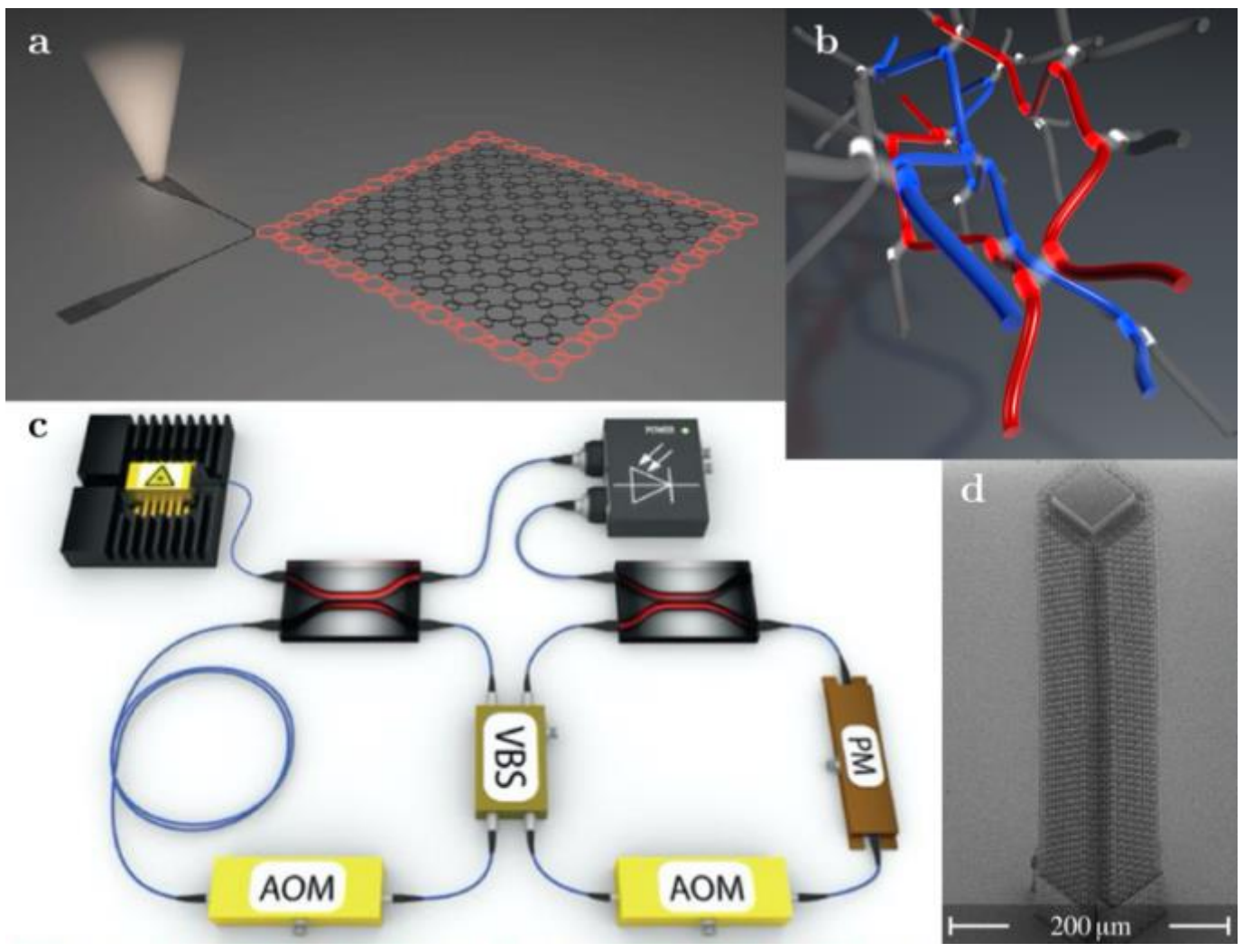

Figure 2 **Future Challenges of Floquet Topological Photonics**. **a,** A topological laser is established by a pumped unidirectional edge mode of a topological lattice. **b,** Complex lattice structures formed by three-dimensional waveguide trajectories in order to realize a photonic Z2 topological insulator. **c,** A fiber loop setup enables Floquet lattices with a huge propagation length. **d,** Polymer waveguides provides a suitable fundament for miniaturisation and the study of coupled higher order modes. (Figures are adapted and reprinted with permission from [7-10])

## Concluding Remarks

In the recent decade FTIs were demonstrated and investigated in various ways. A plethora of novel, complex, and highly applicable phenomena, demonstrated the incredible impact of FTIs in photonics. Going forward, the interplay with the nonlinear and the quantum-optical regimes will be among the principal challenges in this field. Likewise, the connection to industrial applications has to be further developed in the next years by leveraging the numerous promising starting points that have been identified thus far. Clearly, FTIs will play a two-fold role: On one hand, applied photonics research will continue to harness their unique features. On the other hand, FTIs will continue to provide general insights into the fundamental physics of topology. In this vein, FTIs will not only continue to accompany the photonic research in future, but they will be instrumental in shaping it.

## Acknowledgements

A.S. gratefully acknowledges financial support from the Deutsche Forschungsgemeinschaft (grants SZ 276/9-1, SZ 276/19-1, SZ 276/20-1, BL 574/13-1) and the Alfried Krupp von Bohlen und Halbach Foundation.

## 03– Topological pumps and quasicrystals

*Oded Zilberberg*

Institute for Theoretical Physics, ETH Zurich, 8093 Zürich, Switzerland

### Status

A pump consumes energy to perform work that moves a fluid. In topological pumps, an adiabatic and cyclic variation of a Hamiltonian potential moves particles a quantized and robust distance per pump cycle. Such a quantization is ideal for industrial applications, such as variable rate feeders, low-current electronics, slow-light, and spintronics, as well as a metrology standard [1-3]. Furthermore, topological pumps exhibit topological boundary effects that are potentially-useful as spectrally-protected multiplexors for optical classical and quantum gates [4], cf. contribution on Quantum optical effects in topological photonics.

Topology is a global property of a system, corresponding to an integral over local geometrical properties. In topological pumps, a spatial competition between length scales generates multiple spectral bands. These bands exhibit a nonvanishing curvature in response to the time-dependent modulation of the model, where the latter can be thought of as a synthetic dimension [see contrib. on Synthetic dimensions in photonics]. Averaging this time-local geometric property over the whole pump cycle leads to quantized transport; see Fig. 1 for an example.

Due to their potential applications, pumps were widely-explored in electronic mesoscopic systems, see e.g., [2, 3] and citations within. However, as these systems are inherently open to dissipation channels and diabatic corrections, quantization of the transport is challenging to obtain. As a result, only a few years ago, quantized topological pumping was observed for the first time in more controllable cold atoms experiments [7]. Similarly, only recent developments in photonic metamaterials allow for direct access to the topological boundary modes of the pump using input-output experiments [1, 6].

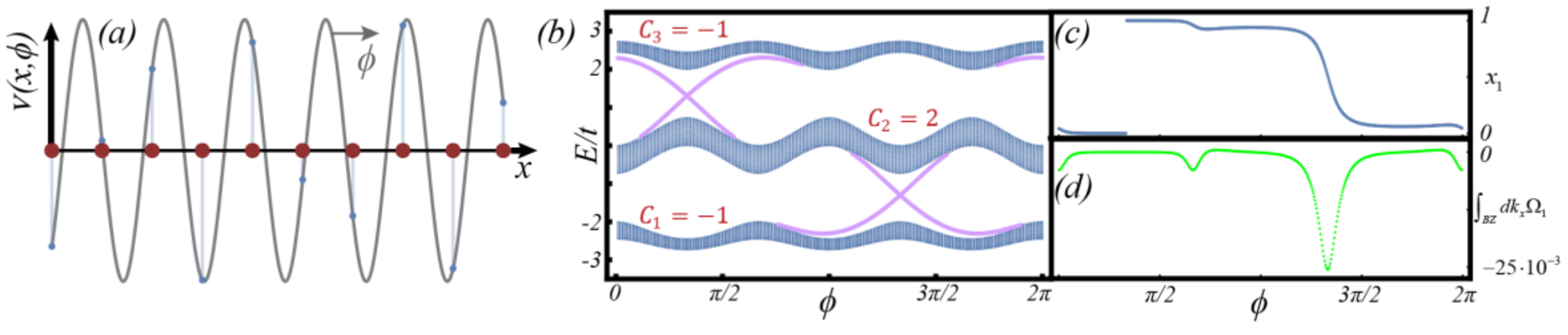

**Figure 1.** *An example of a 1D tight-binding charge pump [5, 6]. (a) Illustration of the model: particles hop between lattice sites (red circles), while sampling* an on-site potential $V(x,\phi)=\cos(2\pi bx+\phi)$ that is cyclic with respect to the pump parameter (a.k.a. phason or synthetic dimension), $V(x,\phi)=V(x,\phi+2\pi)$. The modulation frequency $b$ encodes the length scale competition between the lattice and the potential. (b) Correspondingly, when $b=p/q$ is a rational number (here $b=2/3$), we observe $q$ spectral bands, and when $b$ is irrational, the model can be used to study the topological properties of quasicrystals [6]. (c) Each band $n$ with energy $E_n$ contributes to pumping through the change in its centre of mass $\dot{x}_n=\int_{BZ} dk_x\,\Omega_n\partial_t\phi$, where the integral is over the Brillouin zone of the chain, and $\Omega_n = i\langle\partial_\phi u_n|\partial_{k_x}u_n\rangle - i\langle\partial_{k_x}u_n|\partial_\phi u_n\rangle$ is the Berry curvature associated with the instantaneous Bloch states $u_n$ of the band. (d) The centre of mass motion is quantized with the topological index (Chern number) of the band $C_n=\int_{BZ} dk_x\int_0^{2\pi} d\phi\,\Omega_n$. Correspondingly, at an open boundary, subgap modes thread the gaps as a function of $\phi$ [6], as seen as purple dots in (b).

## Current and Future Challenges

Similar to many other topological effects, topological pumps are best understood within a linear, adiabatic, and dissipationless limit. Contemporary challenges are four-fold and involve (i) extending the zoo of pump models and associated pumped quantities within the linear domain, e.g., by exploring topological pumping and their boundary phenomena in higher-dimensions [1, 8], or in new geometries that transfer quantities other than charge, such as spin or even parity [1]; (ii) including realistic corrections such as non-Hermitian and diabatic effects to the pump [1, 9]; (iii) moving to many-body topological pumps, by considering weak and strong nonlinearities [1]; and (iv) harnessing topological pumps for true industrial applications.

We leave detailed discussion of the plethora of effects in (i) to longer reviews [1]. Similarly, the promise of new applications in (iv) is omnipresent within the contemporary study of topological effects. Yet, a clear technological advantage is still missing. In the context of (ii), which is explicitly relevant to topological photonics, diabatic corrections to topological pumping tend to destroy the topological character alongside the quantization of the pumped charge [2]. Furthermore, loss of particles due to dissipation leads to reduced transmitted power and fidelity of the pump for communication applications. Recently, in an experiment using coupled plasmonic waveguides, quantized pumping in a diabatic system was obtained using a time-dependent modulation of the dissipation. The latter is realized by changing the widths of the waveguides along the propagation axis of the plasmons. This experiment exemplifies the challenges and opportunities in (ii), where non-Hermitian effects are harnessed to obtain quantized pumping.

Moving to many-body pumps (iii), a wide variety of directions opens up in both fermionic and bosonic realizations. Considering more traditional electronic realizations of pumps, weakly-interacting systems are well understood as long as a non-degenerate ground state evolves along the pump-cycle without closing the many-body gap. At the same time, as topological pumps are intimately related to quantum Hall systems, where interesting fractional topological pumps are predicted to appear in strongly-interacting systems [10]. Moving to topological photonics and driven nonlinear photonic systems, the slow modulation of bifurcated steady-states can lead to novel pumping paradigms, where a new topological classification will be required.

## Advances in Science and Technology to Meet Challenges

Technology is racing forward to couple linear and nonlinear elements in controllable quantum engineered systems. The controllability of such devices will allow us to realize the various unexplored scenarios described above. With tunable interactions in cold-atom setups, various strongly-correlated scenarios are within reach, where new topological pumping can occur. At the same time, the lifetime of excitations in such experiments is relatively short, such that quasi-adiabatic topological pumping will be challenging. The combined effort of inventing quantized diabatic pumps that can function at short times alongside constant development of longer-lived coherent platforms will push towards realization of more exotic pumps.

Similarly, nonlinear optics devices with additional slow pumping can lead to new methods to guide light. Here, much remains to be understood concerning the required coherence for such protocols, their topological characterization, as well as, their potential for applications.

## Concluding Remarks

Topological pumps have only recently been experimentally realized, thus opening up a very active research topic, where many-body effects, time-dependent drives, and dissipation interplay to realize new topological effects in the bulk and at the surface of electronics and photonics systems.

**Acknowledgements**
We acknowledge financial support from the Swiss National Science Foundation through grantsPP00P2P2_1163818 and PP00P2_190078. I would like to thank H. M. Price for fruitful discussions.

## 04 - 3D Topological Photonics

*Yihao Yang and Baile Zhang*

Nanyang Technological University

### Status

Three-dimensional (3D) photonic crystals and artificial structures (metamaterials) have played an extremely important role in the history of photonics research (Fig. 1a). For example, it is the pursuit of a realistic 3D photonic bandgap in a 3D photonic crystal that stimulated, in the complex interchange between experiment and theory, the birth of photonic crystals as a modern photonics field [1]. A representative invention is the Yablonovite crystal, named after its discoverer Eli Yablonovitch. The 3D bulk metamaterials not only justify the effective constitution of negative refractive index, a hypothesis of Victor Veselago (1929-2018) back to 1967 [2], but also account for the amplification of evanescent waves that underpins the mechanism of superlens, as proposed by John Pendry in 2000 [3]. It is the 3D geometries that give rise to relatively richer and more interesting physics in the history of photonics research.

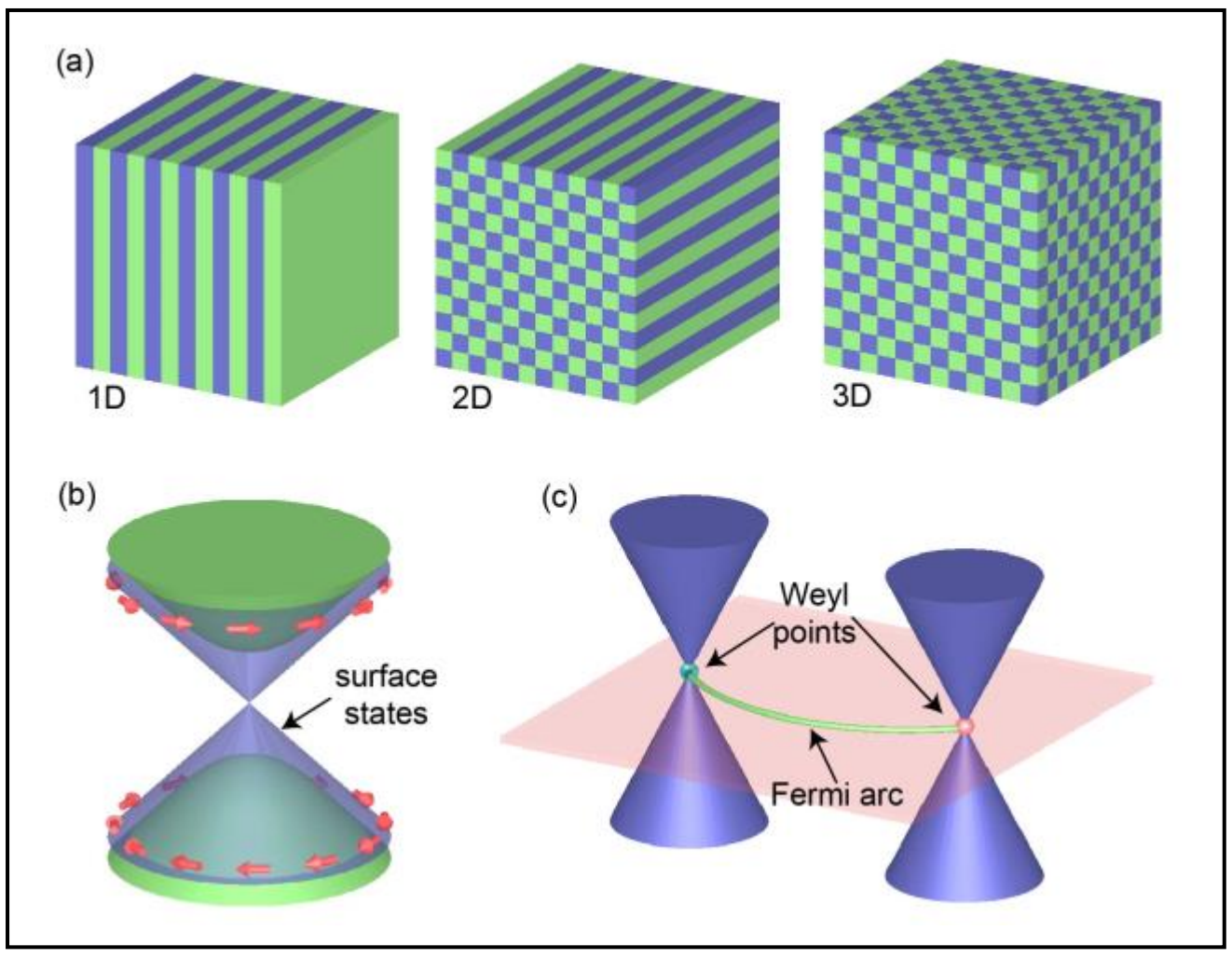

**Figure 1.** (a) Photonic crystals and metamaterials from 1D, 2D, to 3D. (b) A 3D photonic topological insulator is an electromagnetic analogue of a 3D electronic topological insulator. It supports topological surface states forming Dirac cones. (c) Photonic Weyl points, similar to those in 3D Weyl semimetals, carry topological charges in momentum space, being connected with Fermi-arc surface states.

Coming to the new era, we have witnessed an explosive search in novel topological materials, which followed the line from the one-dimensional (1D) Su-Schrieffer–Heeger chain, to the two-dimensional (2D) quantum Hall effect and quantum spin Hall effect (or 2D topological insulators), and lately moved to the climax of 3D topological insulators and 3D topological semimetals. Following the historical tradition of transferring concepts from condensed matter physics to photonics (e.g., photonic crystals function as “semiconductors of light”), there is clearly a necessity of developing 3D topological photonic materials. Currently, 3D photonic crystals and metamaterials are the major platforms to construct photonic analogues of 3D topological insulators and topological semimetals. In particular, a 3D photonic topological insulator shall host in a 3D bandgap topological surface states that form Dirac cones (Fig. 1b), and a 3D photonic analogue of topological semimetals shall carry degeneracies such as Weyl points, which are connected in momentum space by an open Fermi arc (Fig. 1c).

Rather than simply mimicking condensed-matter phenomena, 3D topological photonics may reform some fundamental concepts in photonics, such as negative refraction and spontaneous emission in a cavity, and lead to unprecedented photonic applications. Before that, sufficient platforms of photonic crystals and metamaterials need to be developed. However, current studies in 3D topological photonics still remain relatively rare, due to the challenges stemming from the fundamental differences between photons and electrons.

**Current and Future Challenges**

A typical approach of transferring condensed matter concepts to photonics is to construct photonic analogues of the Schrodinger equation. This works well in 2D but has difficulties in 3D. First of all, light is a vectorial wave, not a scalar electron-like wavefunction. In a 2D geometry, two orthogonal polarizations of light can be decoupled, with each described with a scalar function. This is why 2D analogues of the Schrodinger equation usually work well. However, it is not the case in 3D because of the coupling between polarizations. Secondly, light is volumetric, being intrinsically 3D. Take the typical 2D material graphene for example. The electrons will be confined in graphene without jumping out into the external environment because of the finite work function. However, light guided by a 2D slab can still perceive its surrounding environment with evanescent fields and suffer from scattering by obstacles in a distance from the slab.

Another challenge comes from the missing electron-like spin-up/spin-down states as well as the consequent spin-orbit coupling as in topological materials. Fortunately, several approaches have been developed previously to construct photonic pseudo-spins that behave like electronic spins. A typical example is the superposition of transverse electric (TE) and transverse magnetic (TM) modes [4]. This usually requires the relative permittivity being equal to the relative permeability ($\varepsilon=\mu$). To mimic spin-orbit coupling, bianisotropy (or magnetoelectric coupling) is introduced, which takes effect as synthetic magnetic fields on different pseudo-spins with opposite signs [4]. However, the bianisotropy properties are generally weak in most photonic materials, leading to insufficient bandgap sizes in time-reversal symmetric 3D photonic topological insulators (e.g., the relative bandwidth is only about 1% in the initial proposal [5]). Such small bandgaps severely hinder experimental realization and characterizations.

One may also construct 3D photonic topological insulators with broken time-reversal symmetry, without relying on the pseudo-spins. Along this line of thought, a 3D photonic topological insulator featured with a single surface Dirac cone that is protected by the underlying lattice symmetry was proposed [6]. However, it requires complicated nonuniform magnetization of gyromagnetic materials that are considerably challenging in experiment.

Moreover, the 3D topological phases are not necessarily limited to the photonic crystals in 3D spatial dimensions. They can also be realized in coupled waveguides, continuous media, 3D network systems, and even in synthetic dimensions formed by frequency and geometry parameters. However, experimental efforts toward these directions are relatively few.

**Advances in Science and Technology to Meet Challenges**

Considering the complexity of magnetization, realizing a 3D photonic topological insulator with time reversal symmetry is easier than that with broken time reversal symmetry. It is well known in the metamaterials community that, split ring resonators (SRRs), a classic type of electromagnetic artificial atoms, or “meta-atoms”, exhibit strong bianisotropy. With this insight, a 3D photonic topological insulator consisting of SRRs was experimentally realized in the microwave regime very recently [7] (Figs. 2a-b). Owing to the giant bianisotropy produced by SRRs, the resulting 3D topological bandgap is significantly large (relative bandwidth ≈ 25%), even exceeding the previously demonstrated 2D counterparts (the relative bandwidths are generally less than 10%). Though the experimental demonstration was carried out at microwave frequencies, the design principles should be generalizable to higher frequency regimes, such as terahertz and optical, with the help of advanced micro- or nano-fabrication technology, which has shown the promising 3D SRR fabrication in past years. A potential concern is that metallic losses may destroy the 3D topological states at high frequencies. A possible solution is to employ gain media to compensate for the dissipative losses.

Though realization of photonic Weyl points is relatively easier than that of 3D photonic topological insulators, the implementation of optical Weyl points is not a simple task. One possible approach is to shrink Lu et al.’s double-gyroid photonic crystal [8] (Figs. 2c-d) to nanometre scales. The state-of-art nanofabrication technology has already enabled the realization of single-gyroid photonic crystals. The second approach is to adopt the woodpile structure hosting Weyl points [9]. Historically, the woodpile structures were used to realize the first 3D photonic bandgap crystal in the optical regime. The third approach is based on optically coupled waveguide arrays. Optical Weyl points based on such an approach have been experimentally demonstrated [10]. Owing to the nature of waveguide arrays, the feature size is much larger than the operating wavelength (the lattice constant is around tens of the wavelength), which may hinder the further integration with other photonic devices.

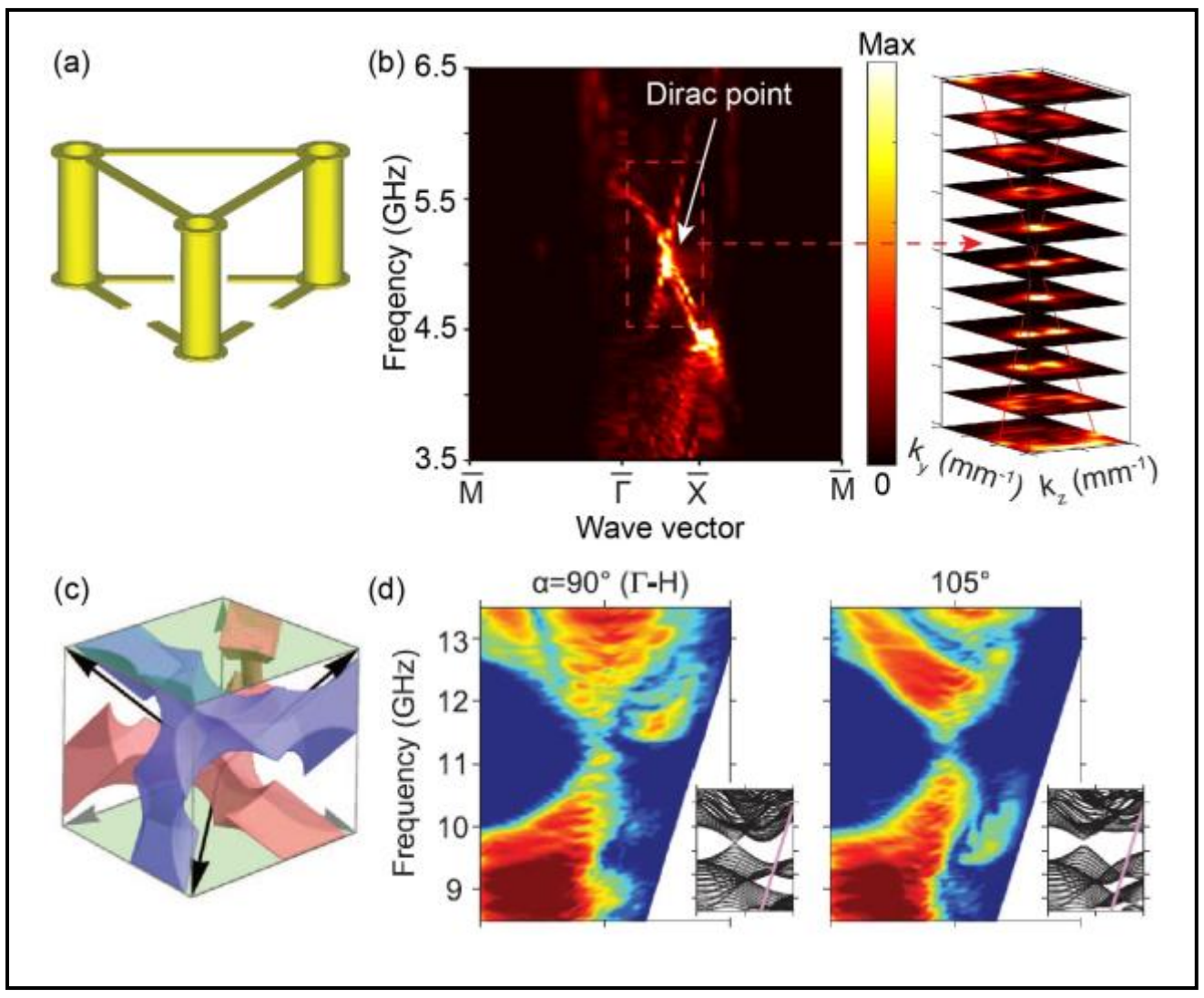

**Figure 2.** (a) Unit cell of the weak 3D photonic topological insulator. (b) Measured Dirac cone of surface states based on microwave field mapping. (a) and (b) excerpted from [7]. Reprinted with permission from Macmillan Publishers Ltd. (c) Unit cell of the Weyl photonic crystal. (d) Projected bulk band structure using angle-resolved far-field measurement. (c) and (d) excerpted from [8]. Reprinted with permission from AAAS.

## Concluding Remarks

3D topological photonics is currently an extremely active and important direction in the entire field of topological photonics. A few breakthroughs, including the realization of photonic Weyl points and a 3D photonic topological insulator, have been achieved in recent years. On the other hand, the development of 3D topological photonics is still in its infancy and future researches will be focused on realization of 3D topological phases at optical frequencies and the further consequent applications. For example, the 3D photonic topological insulators with the ability to confine photons topologically can be used to devise 3D photonic circuitry, cavities, and lasers with topological protection, which may pave a way toward compact robust optical communications and signal processing. The density of states around Weyl points is unique, which may enable novel high-power single-mode lasers. The 3D topological phases would also have important implications for quantum optics and quantum-optical devices, such as interaction between quantum dots and topological surface states, generation of topologically protected entangled photons, and quantum light source that are insensitive to the fabrication imperfection.

**Acknowledgements**

This work is sponsored by Singapore MOE Academic Research Fund Tier 3 Grant MOE2016-T3-1-006 and Tier 2 Grant MOE 2018-T2-1-022(S).

## 05 – Nonlinear Topological Photonics

*Andrea Alù*

CUNY Advanced Science Research Center, USA

### Status

Topological photonics has been providing new degrees of freedom in the quest of manipulating light through engineered nanostructures, offering particularly exciting opportunities to enable robust propagation and resilience to noise and disorder. Given the prominence that nonlinear phenomena have on the functionality of many photonic devices, including lasers, optical sources, switches and modulators, both for classical and quantum applications, it has been a natural question to explore how topological photonics can be combined with nonlinearities to enable robust responses and enhanced light-matter interactions [1]. On one hand, topological phenomena can enhance the response of nonlinear photonic devices and provide robustness to their response, at the same time nonlinear effects can enable reconfigurability and endow topological order in a device based on external optical pumps or even just on the input signal itself interacting with the nonlinearities.

In the past few years, several opportunities arising when topological light phenomena are combined with nonlinearities have been proposed, first theoretically and more recently in a few experimental platforms. It has been predicted that robust soliton propagation can occur in topological photonic crystals within a nonlinear background [2], a concept later experimentally verified using arrays of helicoidal waveguides [3-4]. Signals exciting a nonlinear Su–Schrieffer–Heeger (SSH) array have been shown to trigger a topological transition (Fig. 1a), inducing topological order as they interact with the nonlinearity [5], verified experimentally in an electronic circuit in which the intensity of the input electromagnetic wave has been shown to trigger topological protection to defects [6]. Nonlinear photonic crystals have been shown to support topological order and nonreciprocal responses imparted by suitably tailored external pumps [7] (Fig. 1b). Topological arrays have been shown to provide robust harmonic generation, extending the linear features of topological photonics to inherently nonlinear operations [8]. All these works and several others (see [1] for a review) have provided excitement in this emerging research area, suggesting unique possibilities for basic science and applied technological progress in classical and quantum photonics. Yet, several challenges need to be addressed both from the theoretical and the applied standpoint, as discussed in the following.

### Current and Future Challenges

The challenges that need to be overcome in order to enable a flourishing future for the field of nonlinear topological photonics can be divided into theoretical and experimental nature. From the theoretical standpoint, we should recognize that topological order to date has been inherently rooted into the framework of linear band theory: topological features and corresponding invariants of a photonic crystal are defined in reciprocal space, yet it is not at all trivial how to properly define topological concepts, or even simply band structures, in general nonlinear structures. Given that the field distribution modifies the optical properties through the nonlinearity, we cannot expect the structures to remain periodic after the excitation triggers nonlinearities. So far, these difficulties have been somewhat mitigated through linearization around the local value of the field intensity. In [5], a topological invariant has been defined locally at each point of a periodic array through linearization, and it has been shown that, as the field intensity locally changes, it can close and open bandgaps, inducing a topological transition and a corresponding local switch of topological order (Fig. 1a). This in turn suggests that nonuniform field distributions along a 1D array can enable the generation of moving domain walls and soliton propagation with topologically robust features [9]. While these approaches work in some nonlinear regimes, more work is necessary to properly define topological invariants and the corresponding field evolution in the most general scenarios, and understand the implications of these phenomena, especially in the presence of multi-stabilities.

From the experimental standpoint, nonlinearities in photonic materials are inherently weak, implying that nonlinear topological responses are limited, and require large footprints with stringent phase matching requirements to be observed. Significant efforts have been devoted in recent years to drastically enhance light-matter interactions through hybrid metamaterial platforms, for instance combining multiple quantum wells or 2D materials offering strong nonlinearities, and/or applying advanced photonic concepts, such as Fano resonances and quasi-bound states in the continuum. Combining these efforts to showcase strong nonlinearities with topological features may enhance by orders of magnitude the topological response compared with the state-of-the-art.

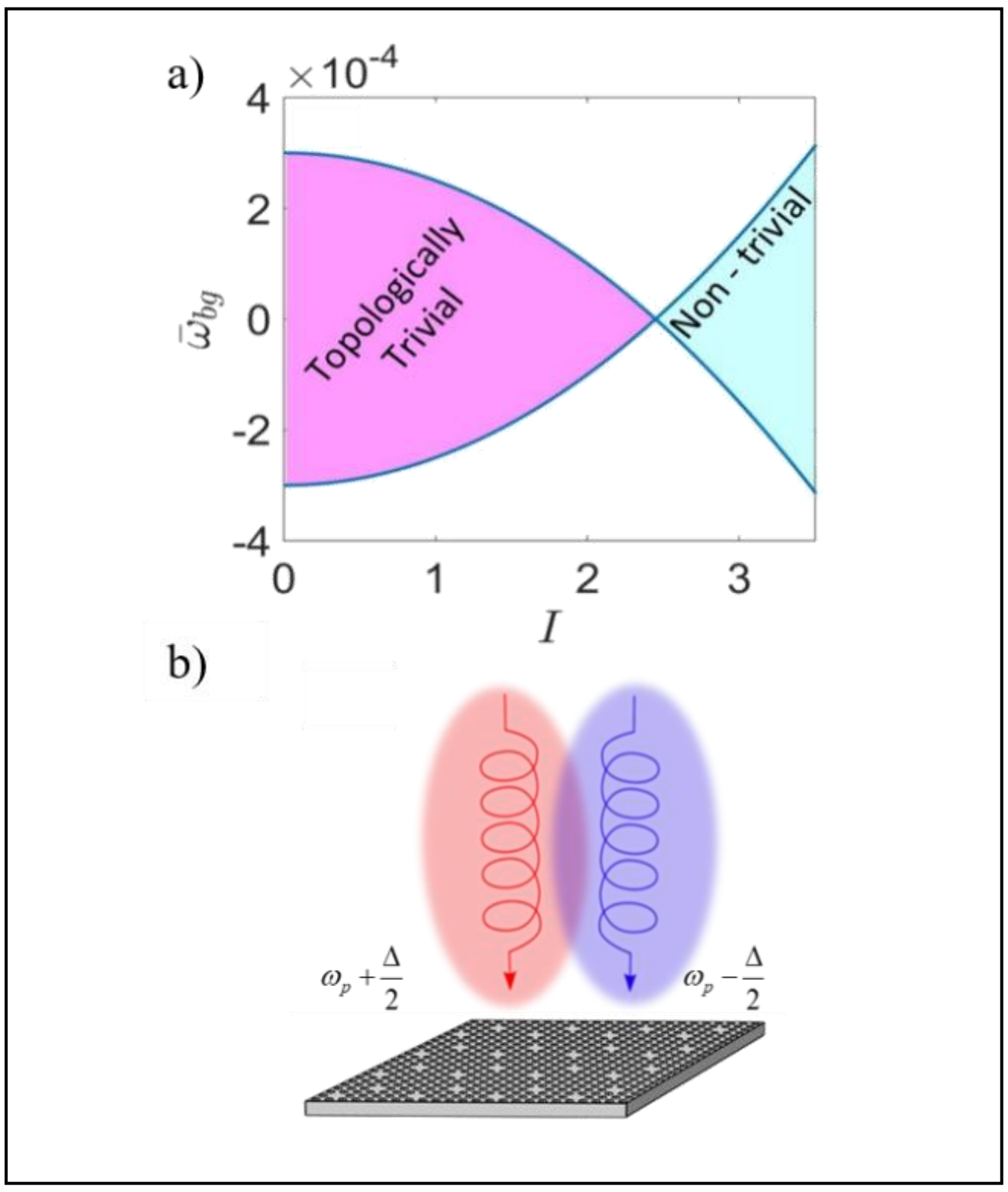

**Figure 1.** (a) Topological transition induced by the varying input intensity in a nonlinear SSH array [5]; (b) A photonic crystal supporting topological order based on optical pumping through nonlinearities [7]

## Advances in Science and Technology to Meet Challenges

Given the mentioned challenges, it is clear that the field of nonlinear topological photonics necessitates a concerted effort to advance the underlying science and technological platform, involving theorists and experimentalists with interdisciplinary capabilities. Fig. 2 shows an example of functionality of a nonlinear photonic crystal [10], whose topological order is controlled and tailored by the same signal exciting the structure, consistent with the topological transition sketched in Fig. 1a. When excited at the top right corner by a low-intensity signal (Fig. 2a), the input signal excites propagation in the bulk of the crystal. However, excitation at the same frequency and location with a higher intensity (Fig. 2b) generates a topologically protected edge mode that travels around corners with immunity to disorder and perturbations. For this functionality to work in a practical and useful nanophotonic device at reasonable intensities, large light-matter interactions are required, which may be enabled exploiting the combination of advanced materials and photonic engineering, as mentioned in the previous section. At the same time, it is important to consider fast material nonlinearities, for instance the timescales of thermal nonlinearities may be too slow to ensure that this functionality is established without unwanted complications and technological challenges. The interplay between the timescales of the nonlinearity and of the involved topological phenomena need to be carefully studied and may actually open added opportunities. Also advanced imaging techniques to observe these phenomena are needed through technological advances, for instance, in near-field imaging techniques that today are mostly limited to low-intensity excitations. Nonlinearities can also enable topological order through external optical pumps. For instance, Ref. [7] showed how a uniform optical pump consisting of two beating signals can endow a photonic crystal with nonreciprocal propagation and strong topological protection. While this solution simplifies the requirements compared to previous demonstrations requiring spatially tailored pump profiles, still its observation requires complex pump-probe setups.

The robust localization of light at the boundaries of topological lattices can actually be exploited in the context of nonlinearities, offering a way to robustly enhance light-matter interactions. In [8] robust light generation at third harmonic has been shown at the edge of a topological array, opening exciting opportunities when combined with the mentioned nonlinearity enhancement mechanisms.

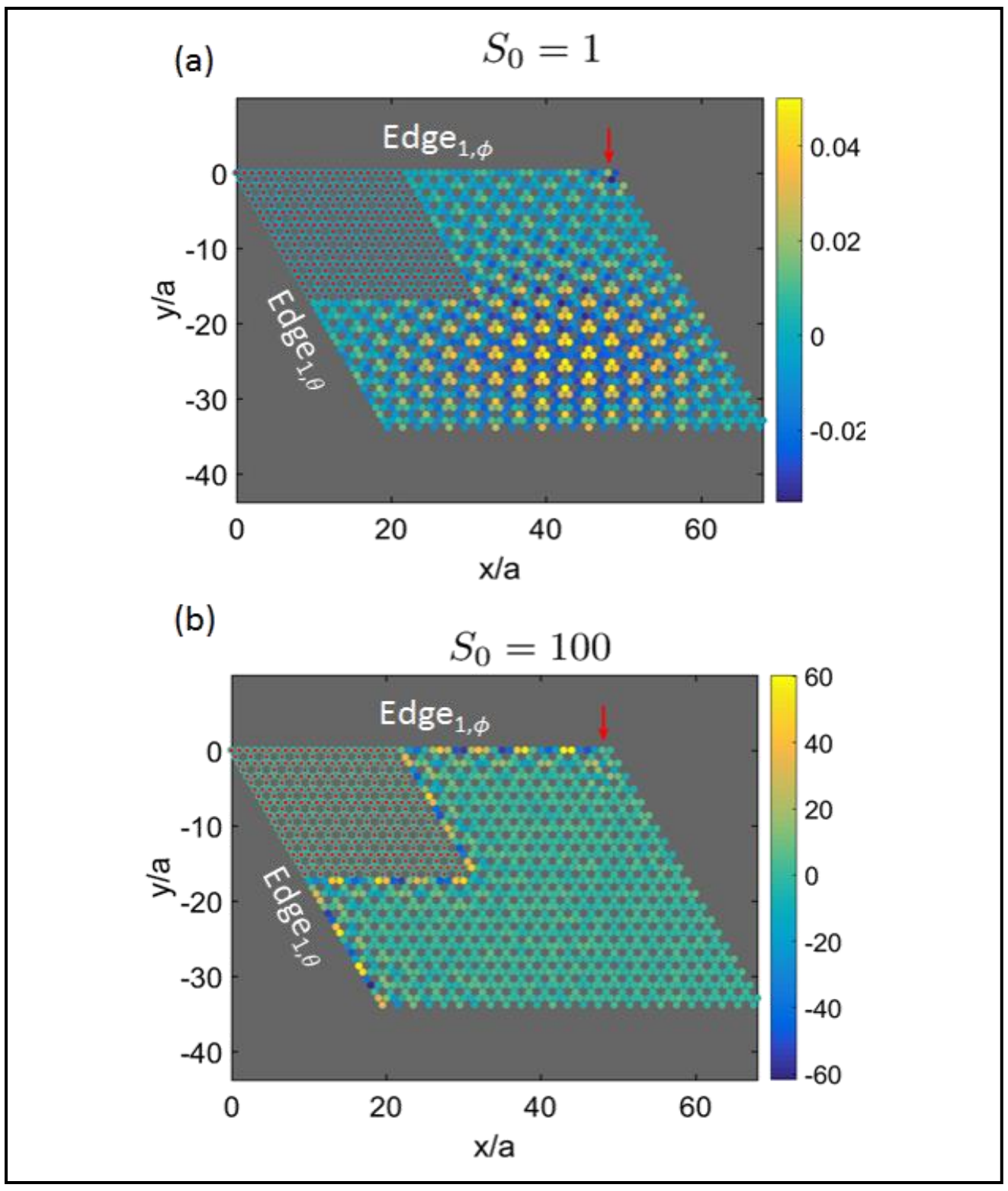

**Figure 2.** A nonlinear photonic crystal whose topological order is changed by the intensity of the input signal [10]. For low intensities (a) the excitation at the upper right corner excites bulk propagation, while for high intensities (b) the input intensity triggers a topological transition, consistent with Fig. 1a, and propagates along the boundary of the photonic crystal.

## Concluding Remarks

The field of topological photonics has been raising significant interest in the past years, due to the fundamental advances that it has unveiled in the control of light, and for the technological opportunities it has opened for optics and photonics. The combination of these concepts with nonlinearities has been emerging as a natural direction for this field of research, offering important opportunities for various devices and applications, yet with several challenges to be overcome both from the theoretical and fundamental science point of view, and in terms of experimental capabilities and applications. Nonlinearity-driven symmetry-breaking, nonreciprocity and topological order have a bright perspective, and we envision a concerted effort from the broad photonics community to establish these concepts in a plethora of devices and implementations, including waveguides, microcavities, electronic circuits, metamaterials. These efforts open unique opportunities for topologically robust and broadband phenomena in the context of optical sources, frequency combs, isolators and multiplexers, switches and modulators, both for classical and quantum light.

## Acknowledgements

Our work in this area has been funded by the Office of Naval Research, the Simons Foundation and the Department of Defense.

## 06 – Non-Hermitian Topological Photonics

*Ronny Thomale*

Julius-Maximilian University of Würzburg

### Status

From the beginning, non-Hermiticity has been a highly relevant issue to be addressed in all photonic systems. This is because any coupling to the environment gives way to potentially dissipative (loss) or accumulating (gain) impact on the photonic system, violating energy conservation or, technically, the Hermiticity of the underlying Hamiltonian. The first attempts to tune gain and loss to achieve a new photonic state of matter rooted in the realisation of parity and time reversal (PT) symmetric photonic systems, even before the concept of topological photonics had been framed [1-3]. It enabled the use of PT symmetric photonics to conceive non-reciprocal light propagation, unidirectional invisibility, and arbitrarily fast state evolution despite limited bandwidth.

Today, the rise of non-Hermitian topological photonics relies on the successful diversification of photonic platforms, allowing for a high degree of tunability of non-Hermitian terms. This includes photonic lattices, waveguides, microcavity arrays, exciton-polaritons, split ring resonators, fibres, and, most recently, topolectric circuits. While loss is manifestly easy to implement yet challenging to avoid, the tuning of gain beyond mere broad-scale lasing is more difficult. Intriguingly, it appears to be the interdependence of non-Hermiticity and additional system properties such as reciprocity which opens way to unprecedented topological phenomena. We are beginning to learn that non-Hermiticity not only quantitatively modifies, but qualitatively expands existing classifications of Hermitian topological systems [4].

Along with analysing how non-Hermiticity as a perturbation affects Hermitian topological photonic states, the field has embarked on new topological phenomena associated with non-Hermiticity: Non-Hermitian degeneracies (exceptional points) allow for enhanced sensitivity of resolving frequency splitting [5], and necessitate a reformulation of topological invariants. The non-Hermitian skin effect, where non-Hermiticity appears due to non-reciprocity, yields an extensive localisation of bulk modes at the edge, challenging the conventional view of topological bulk-boundary correspondence [6]. Exceptional topological insulators, so far only existing as a theoretical concept, establish a new notion of topological universality class due to non-Hermiticity [7]. Overall, the field of non-Hermitian topological photonics is at the verge of accessing uncharted territory.

### Current and Future Challenges

From a conceptional viewpoint, non-Hermitian topology adds significant complexity to the characterisation of a topological state. While it was sufficient for Hermitian states to deduce their topological signature from detecting the topological edge modes alone due to bulk boundary correspondence, it now proves necessary to detect both the non-Hermitian topological edge modes and the bulk. This usually requires excellent control over the entire spectrum, which is difficult to achieve in many platforms. In particular, for non-Hermiticity implied by non-reciprocity, there will be the general task to resolve extensive mode localisation which is associated with multiple lengths scales.

An additional challenge to overcome for any sophisticated non-Hermitian topological state is its dynamical measurability. Energy eigenstates with an advanced, i.e., positive imaginary energy eigenvalue part, make the system unstable, while a state related to an eigenvalue with a large retarded, i.e., negative imaginary part might be damped so quickly that it undergoes the resolution threshold. The topolectrical circuit realization of the non-Hermitian skin effect provides an intriguing proof of principle. In order to resolve all the different localised bulk modes of the non-Hermitian Su-Schrieffer-Heeger circuit, impedance measurements were taken in a range to extend over more than 6 orders of magnitude [8]. This degree of measurability is not always accessible, if possible at all, in most platforms.

In particular, a careful distinction between quantum topology and classical topology becomes increasingly difficult in the face of non-Hermiticity. Since the Berry phase as the source for most topological states of matter known to date [9] relates to parameter space and not phase space, there is in principle a convenient correspondence between classical and quantum topology. This insight is at the heart of understanding why there is such a plethora of classical platforms featuring topological states today, where the only requirement appears to be that the platform be capable of expressing interference. With regard to topological non-Hermiticity, the quantum nature of a given state allows for more diverse channels of dissipation than the classical analogue, e.g. through the dissipation of entanglement via quantum decoherence.

**Advances in Science and Technology to Meet Challenges**

One decisive bottleneck to realise non-Hermitian topological photonic systems is the reliable locally tunable implementation of gain in arbitrary dimensions. Optical fibres, for instance, implement gain through coupling to pulsed laser sources, while topolectric circuits use active circuit elements such as operational amplifiers. As with any power supply, however, this will become more challenging for higher dimensions. At least for those photonic systems that represent networks rather than lattices, one might use projective lower-dimensional overlapping graphs that represent the higher dimensional graph. For this solution, however, as is also the case for any realisation of longer-ranged hopping models underlying a given topological state, most optical platforms need to improve on their ability to go beyond nearest-neighbour coupling. Smart topological engineering might further help to diminish the need for gain implementation. Through the use of non-linearities or refined improvement, one might be able to translate a given active topological system into a passive analogue.

As already alluded to in previous sections, it will prove pivotal to decouple wanted and unwanted dissipation in photonic systems, and to eliminate the latter as much as possible. Any technology transfer of a non-Hermitian topological device will first need to minimise the amount of dissipation, and to only allow it where it is needed to realise the intended topological functionality. This and many other aspects will become more pressing as the community will aim at highly integrated non-Hermitian topological photonic systems.

Furthermore, the technological challenges will naturally depend on the specific non-Hermitian topological functionality one intends to realise. The topological light funnel [10], for instance, uses the non-Hermitian skin effect to accumulate a collective bulk signal at the boundary. Seen as a device, it will require a subtle balance between the accumulation efficiency and loss strength implemented in the system: the more loss the sharper the extensive mode localisation at the boundary, but also the larger the dissipation. It is likely that similar motifs of balanced tuning of dissipation will occur not only for the topological light funnel, but also for other dissipative topological effects yet to be discovered.

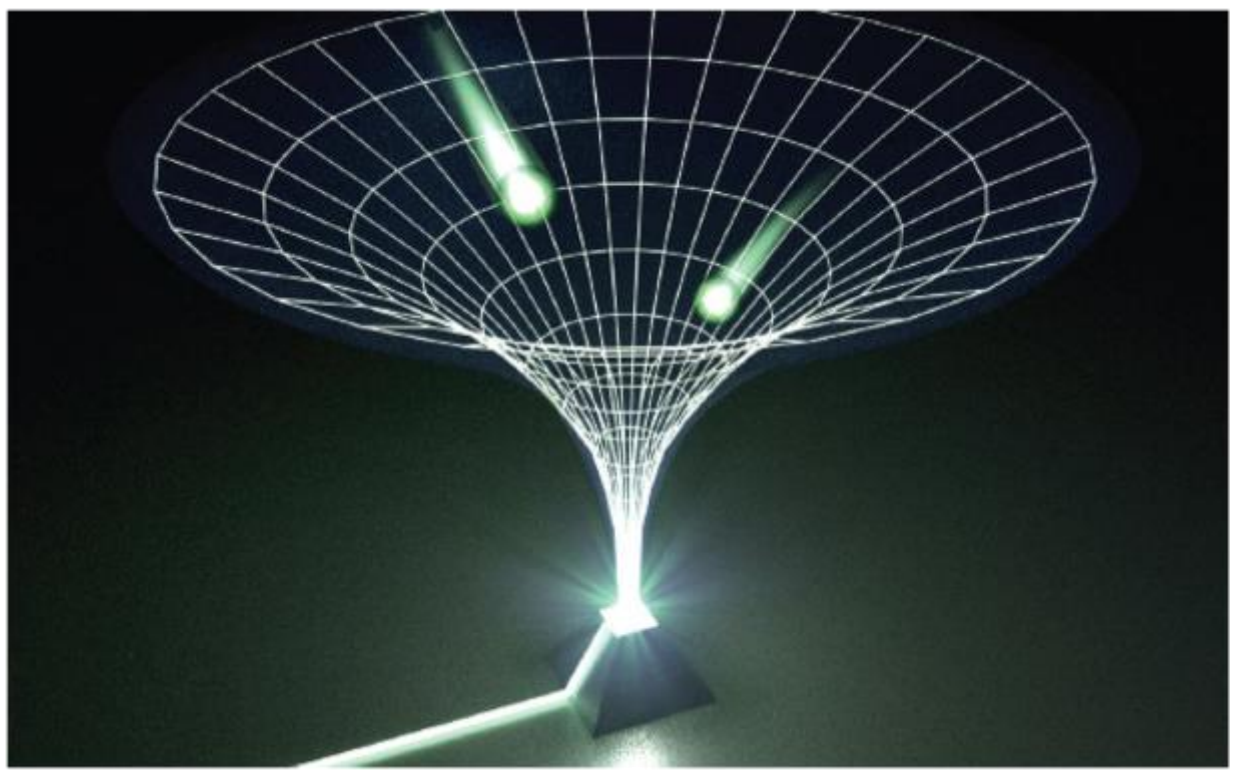

Figure 1. Non-Hermitian photonics might lead to the discovery of new photonic phenomena and unprecedented technological applications. Shown is the schematic plot of a topological light funnel based on the non-Hermitian skin effect (taken from AAAS [10]).

**Concluding Remarks**

A particularly fascinating aspect of non-Hermitian topological photonics is the combined vibrant research activity in theory and experiment. As opposed to Hermitian topological states where condensed matter systems of quantum electrons have been the predecessor to topological photonics, it appears that non-Hermitian topological photonics has the outstanding chance to lead the research effort across all areas of physics. As a whole, the field is too young to tell whether this direction will lead to technological spin-offs in the nearest future. It can be considered likely, however, that groundbreaking progress at the level of fundamental science will unfold very soon, and that non-Hermitian topological photonics will become a pillar for innovation and discovery.

**Acknowledgements**

RT is funded by the Deutsche Forschungsgemeinschaft (DFG, German Research Foundation) through Project-ID 258499086 - SFB 1170 and through the Würzburg-Dresden Cluster of Excellence on Complexity and Topology in Quantum Matter-ct.qmat Project-ID 390858490 - EXC 2147.

## 07 – Strongly interacting topological photons

*Iacopo Carusotto*

INO-CNR BEC Center and Dipartimento di Fisica, Università di Trento, 38123 Povo, Italy.

**Status**

As one can see in the other sections of this Road Map, in most experiments so far the topological features originate from a suitable arrangement of dielectric and magnetic materials; the non-trivial topological invariants displayed by the photonic bands are well predicted by the linear Maxwell equations [1]. Also the nonlinear effects underlying, e.g., topological solitons [2] can be described in a purely wave picture in terms of nonlinear optical polarizabilities such as an intensity-dependent refractive index.

One of the most active axis of research in quantum optics aims at exploring systems whose optical nonlinearities are so strong that non-linear processes are triggered by very weak beams containing only a few photons [3]. The most celebrated such effect is the so-called *photon blockade*, where the presence of a single photon is able to shift the resonance frequency of a single-mode cavity by an amount larger than its linewidth. In this regime, photons can be resonantly injected into the cavity only one at a time and they behave as effectively impenetrable objects. Such *single-photon optical nonlinearities* can be exploited to realize novel optical components, such as optical switches that respond to single photons, with potential applications to quantum information tasks. While the efficiency of basic quantum optical processes may benefit of a surrounding topological environment [4], a really game-changing perspective arises in the many-photon regime of the so-called *quantum fluids of light.*

Light propagating in suitable nonlinear media can in fact be seen as a gas of photons that interact via the optical nonlinearity of the underlying medium. Based on this picture, an impressive series of experiments have demonstrated intriguing many-body phenomena such as Bose-Einstein condensation and superfluidity in weakly interacting photon fluids [5]. Nowadays, the frontier is to combine these studies with single-photon nonlinearities, so to realize strongly (quantum) correlated fluids of light.

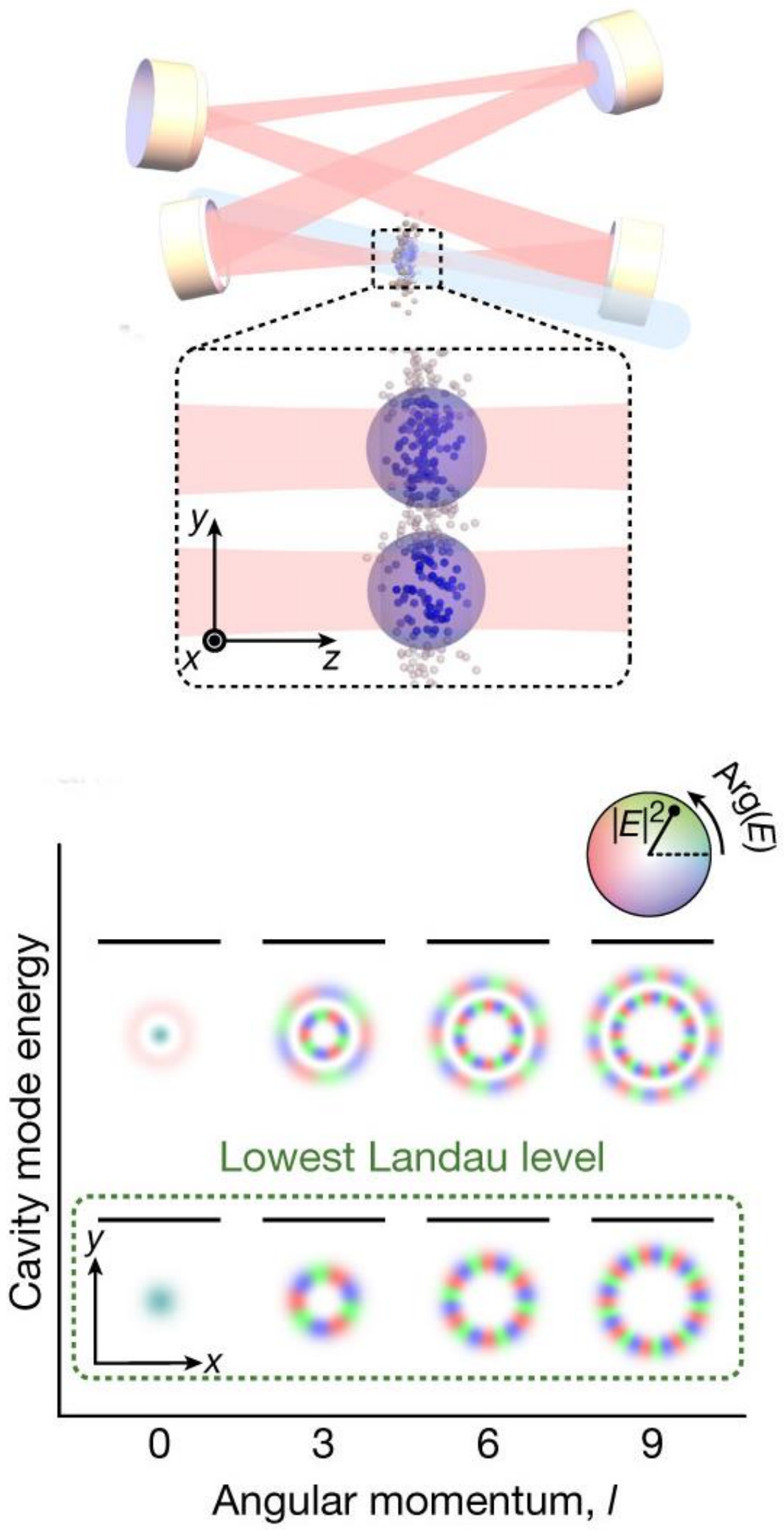

**Figure 1.** Recent experiment demonstrating a two-photon Laughlin state of light. Sketch of the non-planar cavity scheme (top) used to generate the effective magnetic field and, then, the Landau levels for light (bottom) Long-range photon-photon interactions are mediated by dressing with Rydberg atomic excitations (middle). Adapted from [7].

## Current and Future Challenges

After decade-long efforts [5], *Mott insulators* of photons have been recently prepared in 1D arrays of circuit-QED elements [6]. As it happened for cold atom experiments, this is the simplest yet non-trivial strongly interacting state of matter that can be realized with bosonic particles under the effect of strong repulsive interactions. It is characterized by a precisely integer-valued occupation of the lattice sites, with suppressed number fluctuations and a vanishing inter-site coherence.

Even more exciting are the perspectives for 2D systems with non-trivial topological properties. In such geometries, *fractional quantum Hall states* of photons are expected to appear [1] and have been observed in their minimal, two-photon version using the set-up displayed in Fig.1 [7]. In 2D electron gases under strong magnetic fields, fractional quantum Hall states are typically detected by means of electron transport measurements as they are characterized by a fractional (in suitable units) value of the transverse Hall conductivity. In the topological photonic context, more sophisticated probes are required to precisely highlight the nature of the many-body state but these probes will likely offer a deeper information on its features. This includes tomographic insight in the microscopic many-body wavefunction and access to the response to external perturbations, in particular to provide evidence of its fractionalized excitations, the so-called *anyons*. While fractional charges have been observed in shot-noise measurements of the electric current, the fractional statistics of anyons is still awaiting a full experimental characterization. In addition to their conceptual interest as quantum mechanical objects beyond the usual dichotomy between bosons and fermions, non-Abelian anyons were anticipated to offer new strategies to exploit the many-body topology to protect quantum information from noise and decoherence [8]. In this perspective, as compared to ultracold atom or condensed matter systems, electromagnetic and optical systems hold the promise of a straightforward integration into quantum communication networks.

**Advances in Science and Technology to Meet Challenges**

At the moment, one of the main experimental challenges that stands in front of the community is to identify suitable material platforms where such strongly correlated fluids of light can be efficiently realized. Crucial steps in this direction have been reported in many systems, from exciton-polaritons in semiconductor microcavities, to macroscopic optical cavities filled with atomic gases, to optically-controlled spin excitations in ensembles of trapped atoms, to circuit-QED platforms. These latter [9] operate in the microwave window and are naturally endowed with the strong nonlinearities of superconducting Josephson junction elements; they can incorporate strong synthetic magnetic fields so to realize topological models in a rich variety of lattice geometries. As it is illustrated in Fig.2, the rich interplay of these two features was demonstrated in a three-site configuration using a circuit-QED platform [10], but a great difficulty remains to scale up the lattice size. Large lattices are in fact needed to go beyond simple quantum optical effects and host macroscopic fluids displaying a collective many-body dynamics [11].

In parallel to this fabrication challenge, strong efforts are being devoted to the design of efficient protocols for the preparation, manipulation and diagnostics of the fluid of light. Contrary to usual solid-state physics where strongly correlated states of matter naturally appear when the material is cooled down to sufficiently low temperatures, optical systems are intrinsically subject to dissipative processes such as photon losses. On one hand, this raises the problem of efficiently replenishing the photon gas while avoiding undesired heating processes and maintaining the quantum coherence [12-13]. Such fundamental questions are presenting extreme technical difficulties also from the theoretical point of view, as they require new tools for the theoretical modeling of the driven-dissipative many-body dynamics. On the other hand, the intrinsic coupling to the external world widens the artillery of tools that –together with the different forms of multi-particle interactions that can be engineered in these systems– are available to stabilize the desired many-body state and then to manipulate and probe it. The recent advances in the quantum optical tools offer a great flexibility in the design of such pumping and diagnostic schemes, still their practical implementation in the lab may require further technical advances to achieve the required stability and precision.

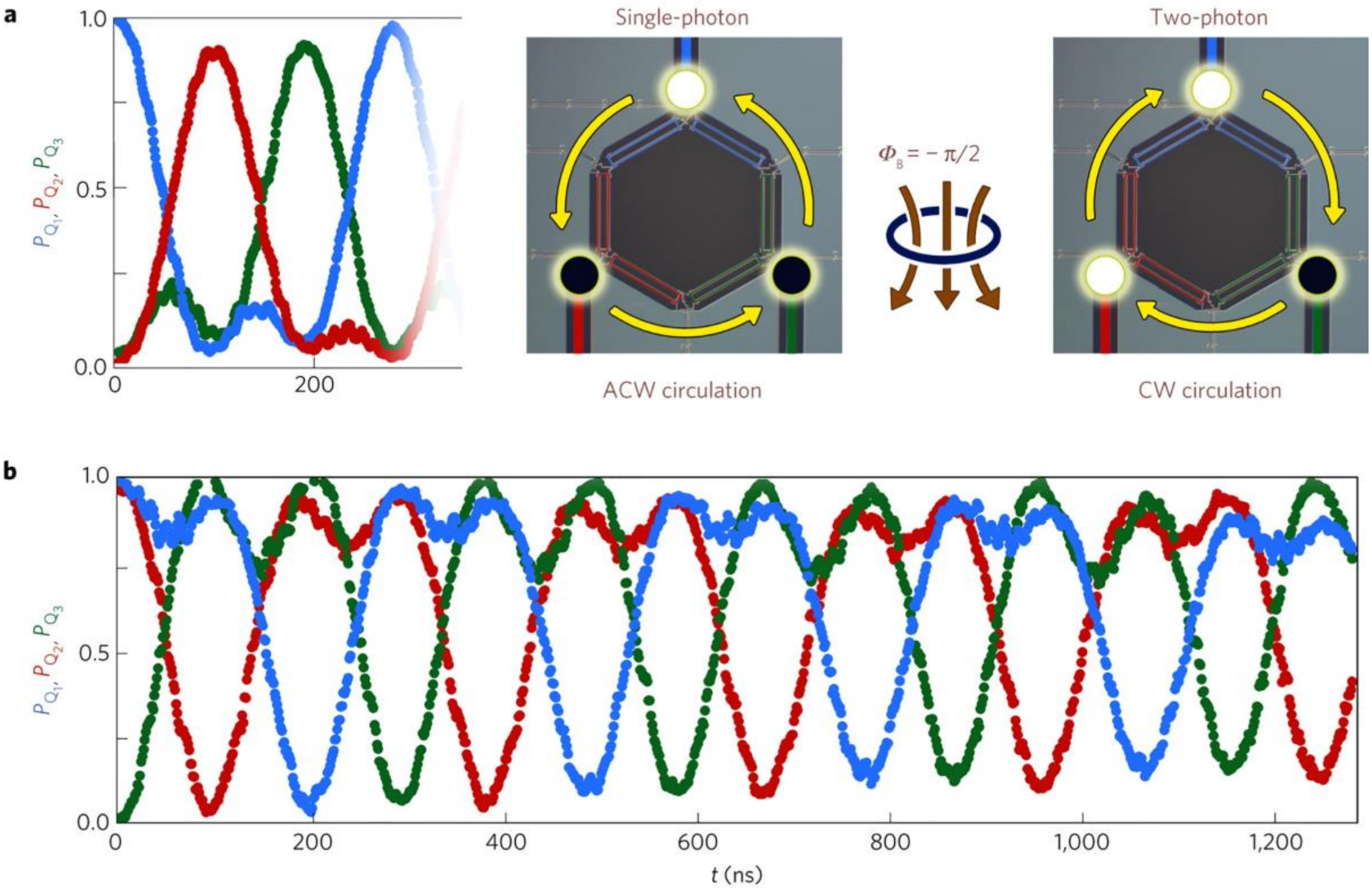

Figure 2. Circuit-QED experiment demonstrating the interplay of strong interactions with a synthetic magnetic field for light in a three-site geometry via the opposite chiral motion of single- (a) an two-photon (b) states. Figure from [10].

## Concluding Remarks

In the last years the research on strongly correlated many body phases in fluids of light has made enormous progress, with the first observation of a Mott insulator of photons, the demonstration of the interplay of strong photon-photon interactions with synthetic magnetic fields, and the first steps towards the realization of macroscopic fractional quantum Hall fluids of light. Much of these advances are based on the inclusion of strong optical nonlinearities into topological photonic systems and offer game-changing perspectives to exploit many-body topology to protect quantum information tasks from external disturbances within optical or microwave quantum communication networks. Given the fast advances displayed by topological photonics in just one decade, I am thus confident that this far-fetched vision will keep offering exciting new developments and raising intriguing questions to a trans-disciplinary community of scientists, from fundamental science to technological applications.

## Acknowledgements

We acknowledge financial support from the European Union FET-Open grant “MIR-BOSE” (n. 737017), from the H2020-FETFLAG-2018-2020 project ”PhoQuS” (n.820392), from the Provincia Autonoma di Trento, from the Q@TN initiative, and from Google via the quantum NISQ award.

## 08– Topological microcavity polaritons

*Philippe St-Jean[1,2] and Alberto Amo[3]*

[1] Center for Nanosciences and Nanotechnologies, University of Paris-Saclay, Palaiseau, France

[2] Université de Montréal, Canada

[3] Laboratory of Physics of Lasers, Atoms and Molecules, University of Lille, France

### Status

The fabrication of materials in which photonic resonances are strongly coupled to electronic excitations has opened the possibility of engineering novel topological properties in the photonics realm. A remarkable example are semiconductor heterostructures in which photons are confined in a Fabry-Perot microcavity tuned in resonance with an excitonic transition of a semiconductor material. If photons are trapped long enough in the cavity to be absorbed by the excitonic transition, emitted, absorbed again and so on, the system is said to be in the strong coupling regime, and the eigenmodes of the system are mixed light-matter quasiparticles called exciton polaritons.

The hybrid nature of these quasiparticles is their main asset for exploring novel linear and nonlinear properties: thanks to their photonic component polaritons can be excited and probed using standard optical techniques, they can be confined in one- and two-dimensional lattices of coupled photonic resonators, and they present polarisation dependent properties which result in intriguing spin-orbit effects; thanks to their matter component, they are sensitive to external magnetic fields causing giant Faraday effects in the near infrared (700-900nm), they show low-threshold lasing, and they present significant polariton-polariton interactions emerging from the coulomb interaction between excitons. These interactions give rise to one of the largest Kerr-type nonlinearities in any dielectric structures [1].

These features have been advantageously employed to explore a number of topological effects in microcavities (see Ref. [2] for a detailed review). For instance, using AlGaAs based structures, a one-dimensional lattice of coupled micropillars was used in the first report of lasing in a topologically protected edge state [3] (see Fig. 1 a-b). Taking advantage of the significant Faraday effect of polaritons in this material, it has been shown that time-reversal symmetry can be broken: the anomalous Hall effect was observed in a planar microcavity [4] and Chern insulating phases with chiral edge states have been implemented in a honeycomb lattice of coupled micropillars [5].

Another very interesting feature of polaritons in a microcavity is the possibility of engineering spin-orbit coupling effects. In a planar microcavity, the polarisation dependent reflectivity inherent to any dielectric interface results in a sizeable splitting between polariton modes linearly polarised parallel and perpendicular to their propagation direction within the cavity. This momentum-dependent splitting can be directly described as a Dresselhaus-like field acting on the polarisation pseudospin of polaritons. When combining it with the polarisation properties of excitons, other field configurations can be achieved. For instance, excitons in monolayer semiconductors like transition metal dichalcogenides or perovskites present very interesting valley-polarisation effects (see Fig. 1 c-d). When embedded in a planar microcavity or combined with a photonic crystal, the strong coupling regime results in modes described by a combination of Rashba and Dresselhaus fields [6]. Even in the absence of excitonic resonances, microcavities embedding liquid crystals have been used to engineer extreme birefringence [7] (see Fig. 1 e-f). These examples show the potentiality of microcavities to engineer exotic forms of spin-orbit coupling which may unveil original topological phenomena.

**Current and Future Challenges**

Three of the most exciting challenges in topological polaritonics are the further development of polariton lattices with broken time reversal symmetry, the study of nonlinear topological phases and the implementation of non-Hermitian effects. As already mentioned, one of the major assets of polaritons compared to other photonic systems is their strong sensitivity to external magnetic fields, even at optical frequencies, which is inherited from the Zeeman splitting of their excitonic component. This allows the possibility of breaking the time reversal symmetry which is a prerequisite for implementing several important topological phases. However, the number of available experimental reports exploiting these feature is extremely reduced [4], [5]. The main reason is that despite their significant value compared to other systems, the observed polariton Zeeman splittings are only of the order of 100-200 μeV at 5-9 T, which is comparable to the typical polariton linewidth, particularly in lattices. This feature has hindered the clear observation of topological gaps induced by the magnetic field.

To tackle this challenge, two complementary approaches are envisioned: improving the polariton linewidth and enhancing the magnetic susceptibility. The former approach will strongly benefit from novel etching and passivation techniques in AlGaAs and InGaAs-based materials, which already have the lowest linewidths. The latter approach could profit from the development of new materials like semiconductors weakly doped with magnetic impurities, like Mn, expected to produce larger Zeeman effects [8]. This dopant might however affect the polariton linewidths through the increase of the material absorption. From a fundamental point of view, the implementation of magnetic effects in polaritonic lattices is of great interest, in particular in combination with the above mentioned exotic spin-orbit coupling effects which may result in original Chern phases. It would also be of interest from a device perspective because it would provide on chip, micrometer-size optical isolation with external magnetic fields low enough to be reached with conventional permanent magnets.

A different area in which polaritons will have a word to say is nonlinear topology. Considering the Kerr-like nonlinearity of polaritons, several theory works have predicted intriguing nonlinear effects for topological edge states, in particular in lattices subject to an external magnetic field. The formation of robust solitons at the edges of topological lattices and their dynamics associated to the presence of the topological gaps are promising research directions. An even more exciting situation is the implementation of topological phases with broken time-reversal symmetry solely emerging from the Kerr-like nonlinearity of polaritons (i.e. without magnetic field). This is possible thanks to the fact that polaritonic lattices can be driven by a laser in resonance with the bands. Starting from a topologically trivial lattice containing point-like band crossings (i.e., a honeycomb or a Kagome lattice), it has been theoretically shown that a topological gap opens in the spectrum of excitations of the polariton fluid when using an excitation laser with a tailored phase pattern [9]. This exotic topological regime has yet to be unveiled experimentally.

Increasing the polariton nonlinearity a step further could push the system into the quantum regime: the presence of a single polariton in a lattice site would induce an energy shift larger than the linewidth. Reaching this regime would open the door to the study of strongly correlated phases, like the fractional quantum Hall effect for photons in the presence of an external magnetic field [10].

Polaritons are also perfectly suited to study lasing in topological landscapes and to engineer non-Hermitian effects. Indeed, microcavities present losses inherent to the escape of photons out of the cavity and the excitonic component of polaritons provides gain. The gain can be locally engineered by pumping the system with an external laser with any preselected spatial pattern. This feature was used to implement lasing in a topological edge state of a one-dimensional lattice [3]. Despite this asset, non-Hermitian features have been hardly addressed using polaritons. This is probably related to the history of the polariton community, which has traditionally tried to extend the polariton lifetimes as much as possible and to reach regimes in which polaritons can be described like a quantum gas close to thermal equilibrium. Although the field of parity-time symmetry and non-Hermitian optics is already quite developed, polaritons can contribute to it by bringing magnetic field effects and nonlinearities in a landscape of suitably engineered gain and losses.

**Advances in Science and Technology to Meet Challenges**

The main difficulty to address the challenges discussed above is the limited lifetime and, consequently, large linewidth of polaritons, in particular in lattices. This is the main obstacle for reaching phases with broken time-reversal symmetry or entering the nonlinear quantum regime. The lifetime is degraded during the etching of the lattices of polariton pillars or mesas, mainly due to the non-radiative traps close to the excitonic transition. More effective passivation of the etched structures or the use of open cavities with patterned external mirrors might be the way to go. A different direction is the use of two-dimensional materials which, as mentioned above, present exotic polarisation properties. Here, the main challenge is the integration in cavities with high quality factors.

**Concluding Remarks**

The hybrid-light matter nature of polaritons provides a unique playground to study topological effects in lattices of resonators. The combination of time-reversal symmetry breaking, significant nonlinearities and exotic spin-orbit coupling effects anticipates the discovery of novel topological photonic phases.

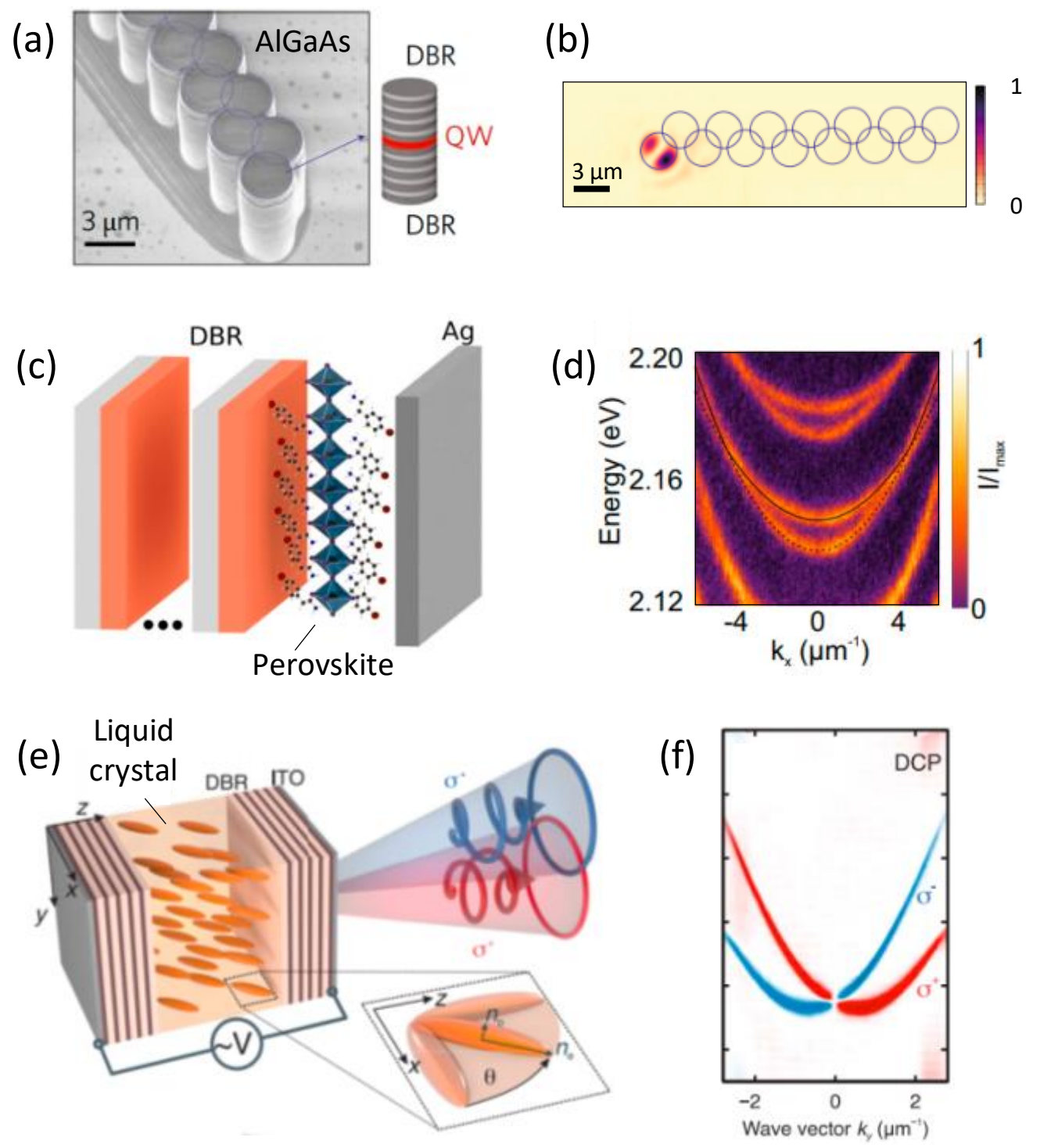

Figure 1. (a) One-dimensional lattice of coupled micropillars. Each micropillars is made of two AlGaAs-based Bragg mirrors (DBR) and a cavity spacer embedding GaAs quantum wells (QW). (b) Measured topological edge mode from the lattice in (a), which implements the Su-Schrieffer-Heeger Hamiltonian. From Ref. [3]. (c) Sketch of a mixed dielectric-metallic microcavity embedding a perovskite crystal. (d) Dispersion from the cavity depicted in (c) showing the crossing of bands with different linear polarisations. From Ref. [6]. (e) Liquid crystal microcavity with giant birefringence. (f) Measured degree of circular polarisation (DCP) in the photonic bands of the microcavity sketched in (e). The applied voltage results in the implementation of a Rashba-Dresselhaus field for the cavity photons. From Ref. [7]. Reprinted with permission from AAAS.

## Acknowledgements

This work was supported by the H2020-FETFLAG project PhoQus (820392), the QUANTERA project Interpol (ANR-QUAN-0003-05), ERC CoG EmergenTopo (865151), the Marie Sklodowska-Curie individual fellowship ToPol, the French National Research Agency project Quantum Fluids of Light (ANR-16-CE30-0021) and CEMPI (ANR-11-LABX-0007), the French government through the Programme Investissement d'Avenir (I-SITE ULNE / ANR-16-IDEX-0004 ULNE) managed by the Agence Nationale de la Recherche and the CPER Photonics for Society P4S.

## 09 - Synthetic dimensions in photonics

*Avik Dutt[1,2], Luqi Yuan[3], and Shanhui Fan[1]*

[1]Stanford University, USA

[2] Department of Mechanical Engineering, and Institute for Physical Science and Technology, University of Maryland, College Park, MD 20742, USA.

[3]Shanghai Jiao Tong University, China

### Status

Topological phenomena are typically richer in systems with more dimensions – an example being the four-dimensional quantum Hall effect, which supports a second Chern number with no lower-dimensional analogue. Additionally, it is of significant interest to realize topological photonics in simpler, experimentally feasible structures, since many demonstrations of two- or higher-dimensional effects relied on arrays of tens to hundreds of waveguides or resonators which are challenging to implement. In this regard, synthetic dimensions have emerged as an attractive concept to explore higher-dimensional physics in simpler, lower-dimensional structures that are easier to fabricate. Here we focus on synthetic dimensions formed by coupling states labelled by an internal photonic degree of freedom, such as frequency [1], [2], spatial/temporal mode structure [3], [4], or orbital angular momentum (OAM) [5] (See Fig. 1). The hopping of photons between these states emulates the propagation of photons along a real-space lattice. After several pioneering theoretical works [1], [2], synthetic dimension lattices have been demonstrated for photons in recent experiments involving transverse spatial modes [3] (Fig. 1(b)), or frequency modes [6], [7] (Fig. 2(d)-(i)).

Three additional benefits of dynamic synthetic dimensions such as frequency or time-bin modes are: (1) long-range coupling and additional synthetic dimensions can be readily engineered since the coupling mechanism is not necessarily linked to the geometric or physical proximity of sites (Fig. 2(g)-(i)), (2) the lattice can be reconfigured, and the couplings between states can be continuously tuned, enabling dynamical studies of topological phase transitions. Both these capabilities [(1) and (2)] are unmatched in real-space lattices. (3) The reconfigurability can be used for actively manipulating internal degrees of freedom (e.g. the spectrum) of light. Further advances in synthetic dimensions are anticipated to realize new phases of light and matter, such as interacting topological phases and higher-order topological insulators. Beyond fundamental explorations, synthetic dimensions also hold promise for applications such as quantum simulation in high dimensions, and photonic machine learning hardware, all in a reconfigurable and scalable fashion.

### Current and Future Challenges

Major research issues and challenges in synthetic dimensions include (1) The creation of boundaries in synthetic space, (2) Extensions to lasers and non-Hermitian lattices, (3) Bringing synthetic dimensions to the quantum regime, (4) The creation of interacting many-body topological phases via nonlinearities, and (5) Miniaturization and on-chip integration for scalability.

While real-space lattices possess a natural boundary, those based on frequency or OAM modes do not possess sharp boundaries. Boundary-related phenomena such as chiral edge states are a hallmark of topological physics due to the bulk-boundary correspondence. The inherent tunability of synthetic dimension couplings could be used to study the emergence of edge effects as the lattice is gradually truncated from an infinite to a finite one. On the other hand, topological lasers, and more generally non-Hermitian photonics, has gained traction in real-space lattices due to the promise of robust, high-power, single-mode lasing, and due to emergence of new physics with no Hermitian analogues, respectively. Synthetic dimensions have much to offer here, with prospects for stable mode-locked pulsed lasers [8], or lasers with tailorable structuring in their transverse or longitudinal mode profiles.

Past research in synthetic space has focused on noninteracting topological phases, i.e. linear systems where a single-particle or classical description is sufficient. A prototypical example is the integer quantum Hall phase [1]–[3], [6] (Fig. 2(a)-(f)). Introducing nonlinearities to implement interacting models would realize more interesting topological phases, such as fractional quantum Hall insulators. Two major challenges towards realizing such interactions are: (i) local interactions in real space are often long-ranged and completely delocalized in synthetic space [9] – an example being the Kerr nonlinearity, which is spatially local but global in frequency space or OAM space since these modes are co-located, (ii) Very strong nonlinearities, ideally at the single-photon level, are required for emulating interacting phases and quantum many-body effects typically studied in condensed matter physics. Such strong local nonlinearities are also important for using synthetic dimensions in quantum computing and quantum simulation. Parametric nonlinearities available in current state-of-the-art photonic materials are orders of magnitude weaker than the required single-photon level nonlinearity. An intermediate near-term goal is to investigate classical nonlinear effects (e.g. soliton-like behavior [10]) on synthetic-dimensional topological photonics.

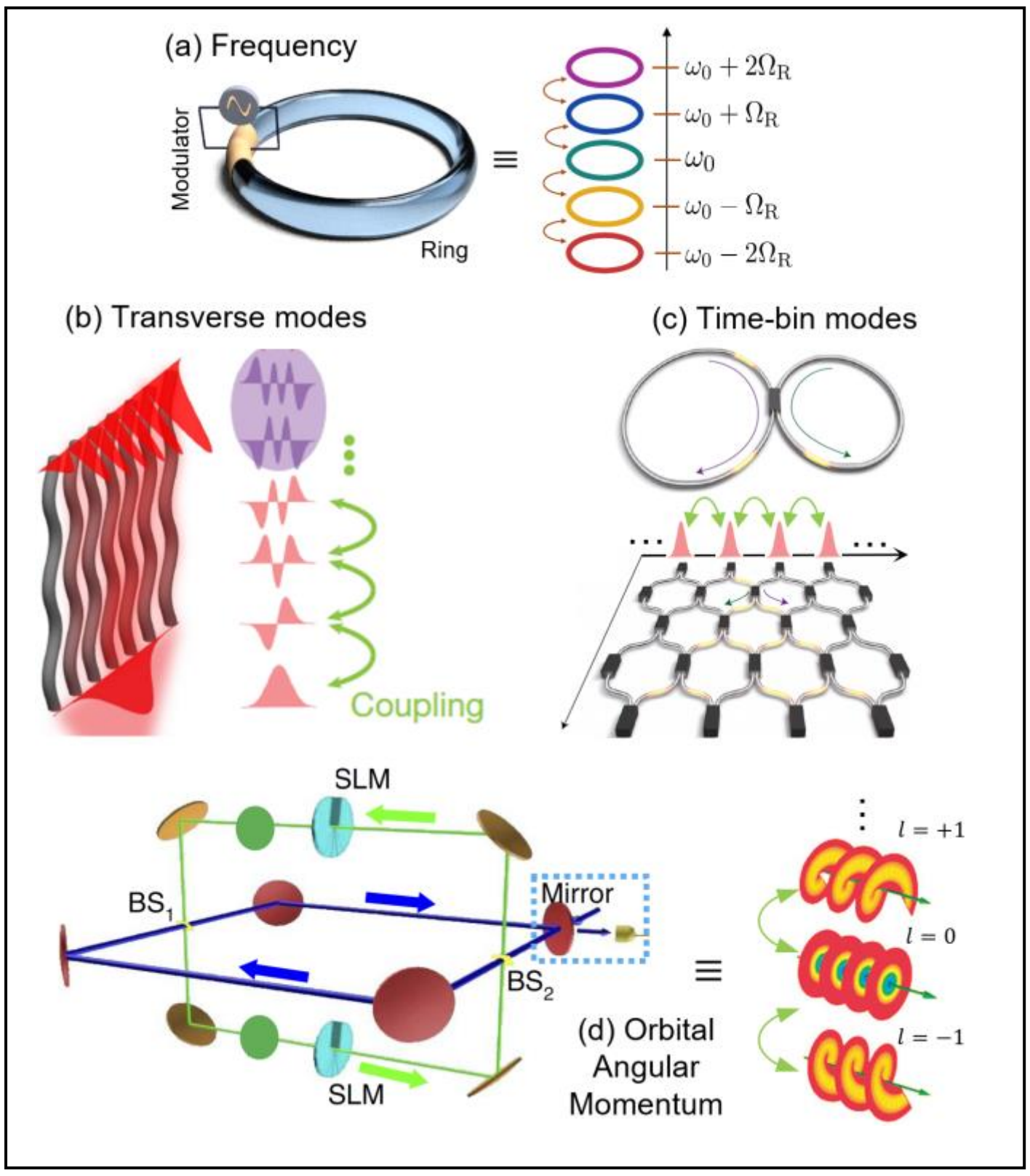

**Figure 1.** Examples of photonic synthetic dimension implementations. (a) longitudinal frequency modes coupled using an electro-optic modulator. Adapted from [1]. CC BY 4.0. (b) Transverse spatial modes coupled by oscillating waveguide arrays. Reproduced with permission from [3]. Copyright 2019 Springer Nature. (c) Time-bin modes in two coupled fiber loops of unequal lengths [4]. (d) Orbital angular momentum (OAM) modes in a cavity coupled using spatial light modulators (SLMs). Adapted from [5]. CC BY 4.0.

## Advances in Science and Technology to Meet Challenges

A combination of strategies such as spatiotemporal structuring, and the introduction of new materials with strong nonlinearities and gain into synthetic dimension platforms is anticipated to address several future challenges. To tackle the lack of boundaries in certain synthetic dimensions, one can either use dispersion engineering (for frequency lattices), or additional spatial structuring (for frequency and OAM lattices) or fast time-varying modulation signals (for time-bin lattices). Synthetic lattices have hitherto been implemented in dielectric photonic structures (waveguides and cavities). Incorporating gain materials and mechanisms to modulate gain and loss in synthetic space is important for topological lasers and non-Hermitian physics. Similarly, by incorporating quantum emitters into dielectric photonic structures, single-photon nonlinearities can be realized, paving the way to interacting topological phases and quantum simulation. In this context, strong coupling of the quantum emitter and photons would be enhanced by on-chip integration due to the small mode volumes achievable.

While initial experiments in synthetic dimensions used large waveguides and resonators [3], [6], recent work in nanophotonic lithium niobate modulators has integrated them on chip and realized multiple frequency dimensions [11]. Further progress towards on-chip synthetic lattices can significantly scale up the number of dimensions, both in real and synthetic space, and provide access to topological phenomena. Particularly for implementing synthetic frequency and time dimensions, modulators in materials such as aluminum nitride, lithium niobite and silicon are attractive. Developments in these platforms are achieving strong modulation at high speeds (>GHz), making them ideal for on-chip synthetic lattices. Progress in materials and technological platforms would also extend synthetic dimensions to platforms beyond optics, including superconducting circuits and optomechanics.

More generally, on the theoretical as well as experimental front, there is significant scope to realize high-dimensional Hilbert spaces using multiple synthetic dimensions simultaneously [12]. The development of protocols that create large entangled states and many-photon cluster states in multiple synthetic dimensions would greatly benefit the scalability of current photonic quantum technologies.

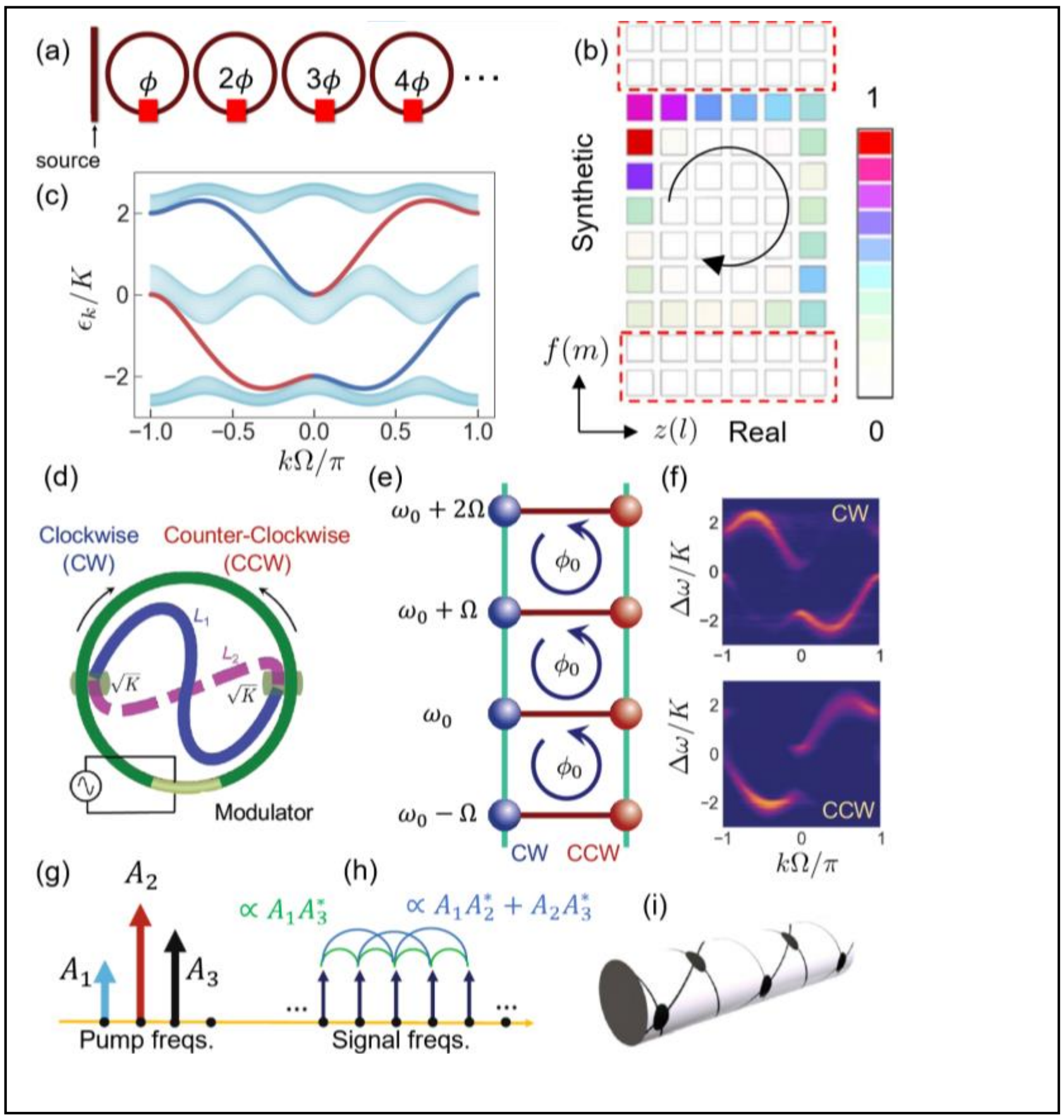

Figure 2. Realizing gauge fields and topological physics in synthetic frequency dimensions. (a) A 1D array of coupled ring resonators, each modulated at the free-spectral range but with a different phase [1]. (b) Chiral one-way edge states in the lattice corresponding to (a) with 1 real dimension and 1 synthetic frequency dimension. Reproduced with permission from [1]. Copyright 2016 OSA Publishing. (c) Band structure of a finite array of rings for $\phi = 2\pi/3$ showing bulk bands (light blue) and edge states localized on the left edge (dark blue) and right edge (red). (d) Experimental realization of the model in (a) using two synthetic dimensions of frequency and spin within a single modulated resonator with an 8-shaped coupler that couples the CW (clockwise) and CCW (counterclockwise) spins [6]. (e) Its corresponding lattice showing a magnetic field piercing the synthetic quantum Hall ladder. (f) Experimentally measured spin-projected band structures directly revealing chiral edge modes of (c). (c), (d), (e) and (f) are reproduced with permission from [6]. Copyright 2020 American Association for the Advancement of Science. (g)-(i) Realization of frequency dimensions using pump-induced nonlinear four-wave mixing, with long-range coupling and multiple frequency dimensions. Reproduced from [7]. CC BY 4.0.

## Concluding Remarks

Synthetic dimensions have emerged theoretically as an attractive concept for exploring topological photonics in high dimensions, yet using simple structures that are easier to implement. While experiments are still in their infancy [3], [6], [11], progress in materials and especially in modulation platforms is bringing several theoretical proposals within the reach of near-term state-of-the-art technologies. On the one hand, incorporating quantum many-body aspects in synthetic dimensions promises to emulate interesting Hamiltonians and has prospects for analog quantum simulation. On the other hand, completely new directions such as topological lasers, non-Hermitian physics, and topologically protected ways of manipulating the spectral, spatial and temporal properties of light are unique to photonics, that break away from the paradigm of condensed matter physics. With these motivations, we anticipate research in synthetic space to provide significant opportunities for both fundamental science and practical applications.

**Acknowledgements**

This work is supported by a Vannevar Bush Faculty Fellowship (Grant No. N00014-17-1-3030) from the U.S. Department of Defense and by MURI grants from the U.S. Air Force Office of Scientific Research (Grant Nos. FA9550-17-1-0002 and FA9550-18-1-0379). L.Y. acknowledges support from the National Natural Science Foundation of China (11974245).

## 10 - Topological bound states in the continuum

*Xuefan Yin and Chao Peng*

Peking University, China

### Status

Bound states in the continuum (BICs) [1] are a class of counter-intuitive states that reside inside the continuous spectrum of extended states but perfectly localize in space and possess infinite lifetimes in theory. As ubiquitous phenomena in many wave systems, the BICs can be interpreted as destructive interference between radiation channels, while their underlying nature is topological [2]: the topological defects in momentum space, represented by integer topological charges carried by polarization vectors. Topological charges are conserved quantities -- they continuously move in momentum space, unless one drops out of the light cone or annihilates with another one with opposite sign.

The topological interpretation of BICs offers a vivid and practical picture in manipulating the radiation of photonic devices such as photonic crystal (PhC). The evolution of topological charges leads to non-trivial observable consequences. For instance, by merging multiple integer topological charges towards the Brillouin-zone centre (Fig. 1a), the scaling rule of radiation in momentum space is modified accordingly, resulting in an effective suppression of out-of-plane scattering losses for record-breaking high quality factors (Qs) [3]. Besides, as shown in Fig. 1b, by gradually breaking structural symmetries -- both vertical symmetry and C2 symmetry, a BIC disappears without symmetry protection and  the integer topological charge carried by the BIC split into two half-integer topological charges carried by circularly polarized states (CPs), keeping moving in momentum space until they bounce into each other. In this case, the combined integer charge only resides at single side of PhC slab, and thus spawn a class of unidirectional guided resonances (UGRs) — resonances that radiate only towards one side of a PhC slab [4].

Since the Qs of BICs can be infinite in theory, they are promising candidates for light trapping — a fundamental functionality desired by numerous applications in photonics, such as lasers, filters and sensors. The ultra-high-Q nature of BICs enables low-threshold lasing [5], making it an ideal laser architecture for on-chip integration. The intricate radiation patterns around BICs [6] also boost the development of vortex micro-lasers [7]. Moreover, recent progress shows that extremely small modal volume and high Qs can be achieved in a nanoscale Mie resonator hosting quasi-BICs, which greatly enhances nonlinear optic processes [8,9].

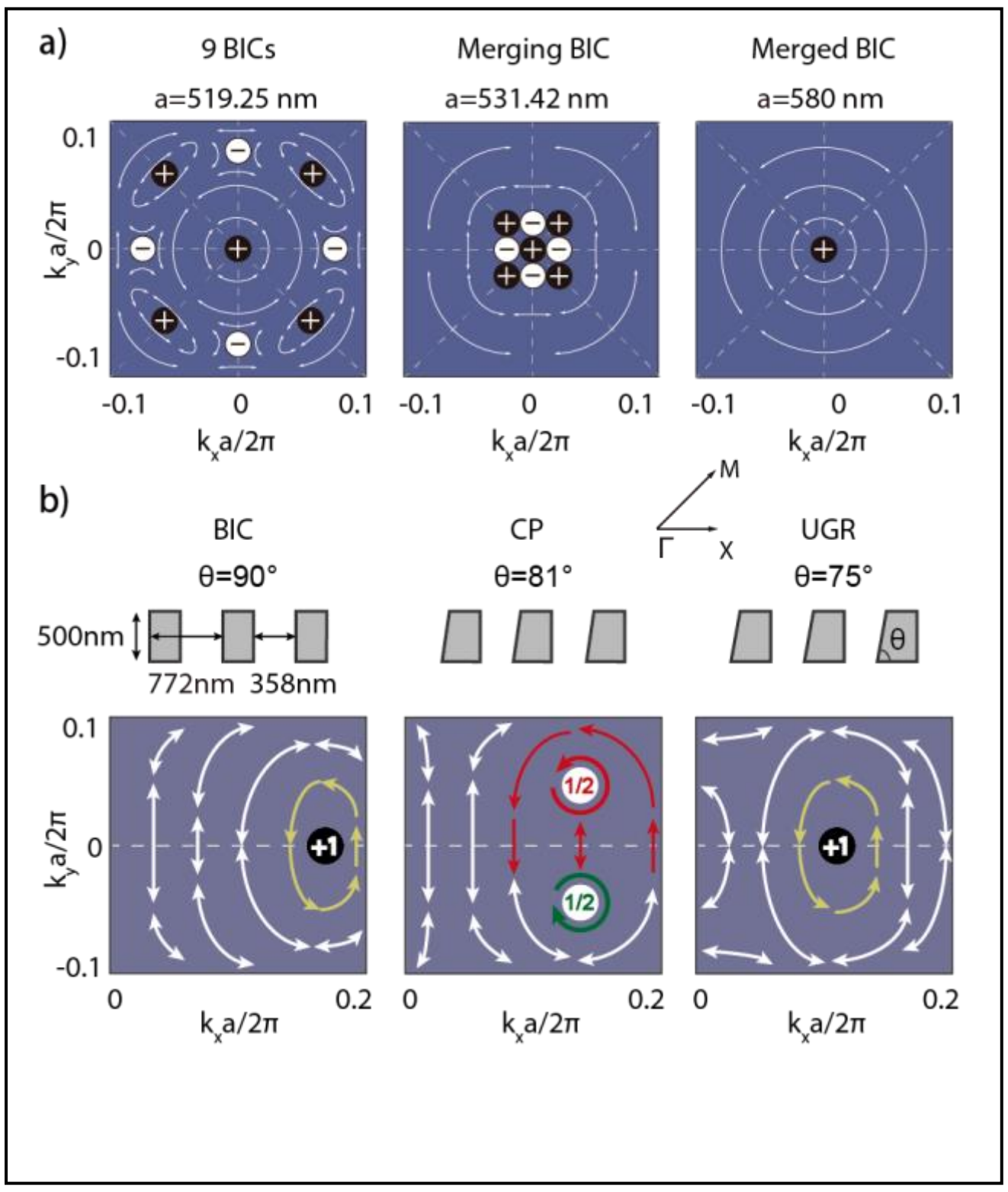

**Figure 1.** Topological charges movement in manipulating the radiation. a, Configuration of charge movements in merging BICs by gradually changing the crystal constant *a*. b, Configuration of charge movements of UGRs by changing the slope of the sidewall. Panel a is reproduced with permission from [3], copyright 2019 Springer Nature; panel b is reproduced with permission from [4], copyright 2020

## Current and Future Challenges

From the perspective of science, other than integer topological charges carried by the BICs, there also exist fractal topological charges. For instance, half-integer topological charges carried by CPs can merge and generate the UGRs [4], but they are actually topological trivial from the perspective of topological band theory. The half charges accompanied with the exceptional points (EPs) (Fig. 2b) are likely not topologically trivial because they reflect the non-trivial Berry phase enclosing the EPs (Fig. 2a) [10]. It is still vague how the integer and fractal topological charges together evolve and interplay in momentum space, and whether they obey a global charge conservation rule, in particular when the system is non-Hermitian. A comprehensive exploration upon the relationship between the "radiation topology" and "band topology" could potentially lead to deeper understanding of the BICs, and connect the topological charges with other branches of topological photonics, such as non-Hermitian topology.

From the perspective of applications, the most challenging and significant tasks for utilizing the BICs are to promote the Qs and shrink the modal volume. Comparing with conventional ultra-high-Q resonators operating outside the continuum and protected by the total reflection, such as micro-rings and micro-disks, BICs rely on the symmetries of the resonators and periodicity of the lattice deeply. As a result, due to the inevitable fabrication error, the realizable Qs of the BICs are still much lower than theory, thus limiting their applications in many aspects. Moreover, unlike conventional optical resonators, the BICs cannot be ideally localized in compact three-dimension (3D) due to the continuity of electromagnetic fields, resulting in relatively large modal volume. For applications desiring small footprint, we look for quasi-BICs [9]: a majority of radiations are cancelled out that guarantee the ultra-high-Q nature, while tiny residual radiations still exist to ensure the non-existence of 3D BICs. In this case, the modal volume could be possibly shrunk to be compact, as shown in Fig. 2c-f. Besides, the BICs are ubiquitous phenomena when the "continuum" is appropriately defined. Given that generality of the BICs, it is interesting and useful to extend the concept to different materials (lithium-niobite, silicon-nitride, and III-V, metal films and etc.), structures (PhC, metasurface, Mie scatters) and wavelengths (ultra-violet, visible to mid-infrared), to make various novel photonic devices.

**Advances in Science and Technology to Meet Challenges**

A comprehensive investigation of topological charges requires the advances in both theory and experiment. For PhC slab, the full-wave coupled-wave-theory (CWT) can be a promising framework since it analytically depicts the radiation and energy band simultaneously [3,4], thus enabling the investigation of the topology of both. For metasurfaces and single resonators in which the wave interactions are short-ranged and dominated by adjacent or one unit-cell(s), the multipolar expansion theory could be a suitable candidate to study their topology [8,9]. When the gain or loss exist in the system, it is expected that the BICs would interplay with the EPs or CPs, revealing the underlying physics of the global topological charge evolution. From the perspective of experiment, the techniques of measuring the radiation polarization had been well established. Although the escaped photons degrade the Qs and induce non-Hermiticity, they also carry the topological information out of the system for the observation.

The challenges of practical usages of BICs or quasi-BICs, including promoting Qs, shrinking modal volumes, generating complex radiation patterns, ultra-fast tunability, and others are solving rapidly through the effort of entire research community. The topological configuration of merging BICs shows a promising way in suppressing random scatterings and resulted in significant higher Qs under inevitable fabrication imperfections [3]. Besides, several exciting advances had been made in achieving high-Q quasi-BICs upon small footprint of metasurfaces [8] and nano-resonators [9], which reveals the great potential of the BICs in nonlinear optics. Moreover, the BICs also enabled an ultra-fast control of vortex micro-laser based on a perovskite metasurface, by which the switching of vortex beam has been demonstrated in 1 to 1.5 picoseconds times-scale [7].

Combining the technology advance in design and fabrication, the BICs will facilitate many applications in which Qs and modal volume are concerned, such as on-chip lasers, nonlinear devices, ultra-sensitive sensors; or utilized as primitive blocks of strong-coupling effects, to build more sophisticated system such as photonic machine learning or quantum computing. we are optimistic that the BICs are no longer fantasy told by the prophets, but are steadily matching forward to realistic applications.

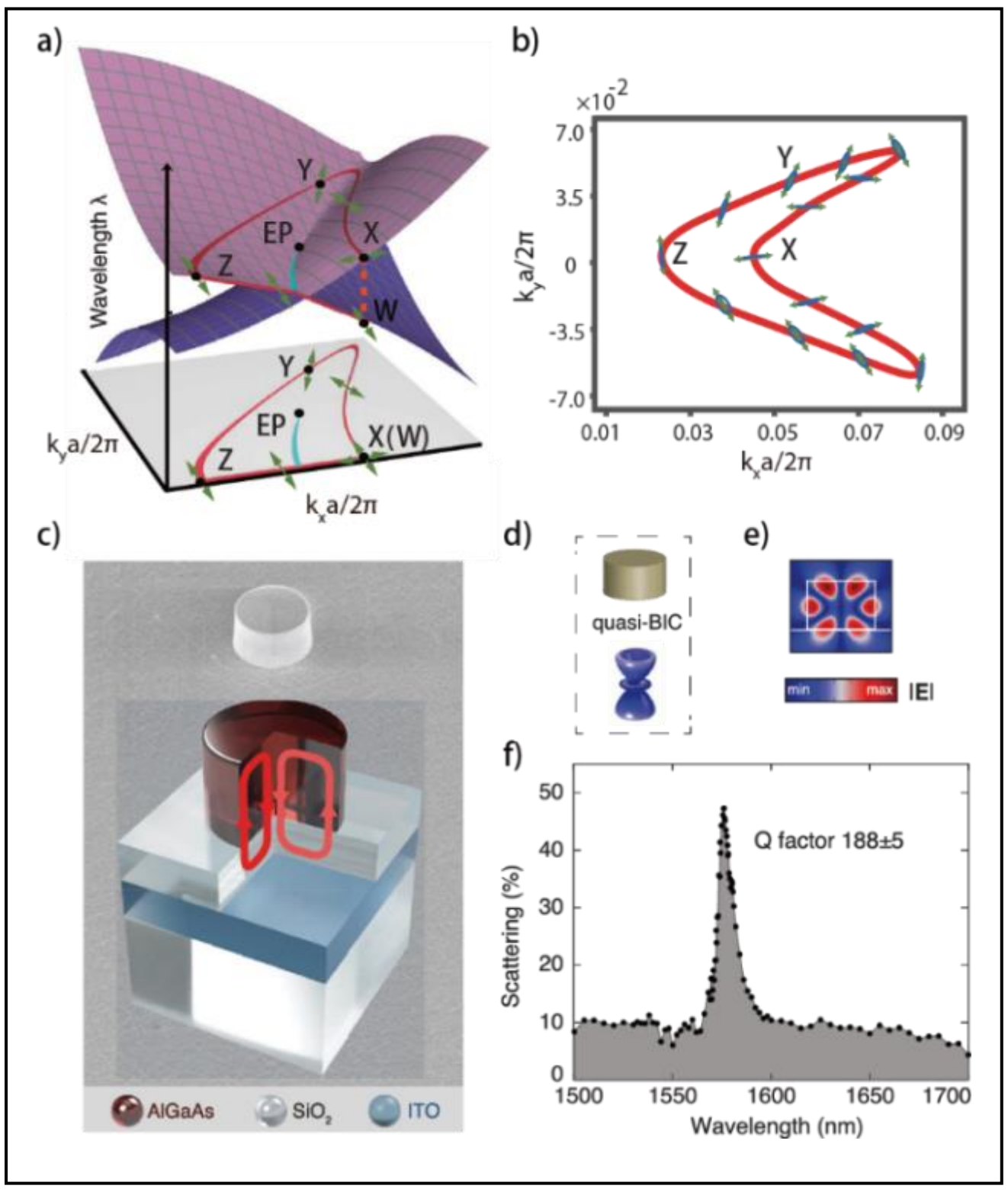

Figure 2. a, Band structure around EP in PhC slab, Z-Y-X gives an enclosed loop around EP. b, half-integer topological charge round the EP. c, Scanning electron micrograph (top) and schematic (bottom) of a nano-resonator supporting the quasi-BIC. d,e Simulated far-field pattern (d) and near-field pattern (e) of the quasi-BIC. f, Measured scattering spectrum of the quasi-BIC. Panels a,b are reproduced with permission from [10], copyright 2018 American Association for the Advancement of Science (AAAS). Panels c,d,f are reproduced with permission from [9], copyright 2020 American Association for the Advancement of Science (AAAS).

## Concluding Remarks

The BICs, a nearly century-old concept from quantum mechanics and a ubiquitous phenomenon that occurs in many fields of wave physics — is finding a new life in photonics. The topological nature of the BICs, namely the integer topological charges, not only broadens the horizon of understanding their origins in physics, but also bridges multiple burgeoning research fields including topological photonics, non-Hermitian physics and singular optics. Besides, the successive advances of BICs upon light trapping reveals their great potentials in photonics and optoelectronics applications of laser physics, photonic integration, nonlinear optics, photonic computation, sensing, and etc., particularly owing to their nature of topology protection. With going deeper into their physics and the advances in design, fabrication and measurement technology, we are looking forward that the obstacles towards practical usages of BICs to be solved eventually, turning the BICs from fantasy to reality.

## Acknowledgements

This work was supported by the National Natural Science Foundation of China under Grant 61922004.

## 11 – Topological Lasers

*Iacopo Carusotto,* INO-CNR BEC Center and Dipartimento di Fisica, Università di Trento, 38123 Povo, Italy

*Tomoki Ozawa*, Advanced Institute for Materials Research, Tohoku University, Sendai 980-8577, Japan

**Status**

Laser oscillation from topological edge states, known as *topological lasing,* is arguably the most distinguishing manifestation of a topological photonic band structure that has no analog in electronic topological insulator materials. Topological lasing can occur either in topological edge states of a Hermitian model where optical gain is externally added, or in edge states of a non-Hermitian model where the gain is included in the bulk model itself.

Topological lasing was first theoretically proposed in 2016 [1,2]. The first experimental realization of topological laser appeared in 2017 in an exciton-polariton lattice where the topological edge state of a one-dimensional Su-Schrieffer-Heeger (SSH) model was made to lase when the gain is spatially located close to an edge [3] [Fig.1(a)]. Soon after, experiments in a 1D chain of microring resonators followed, where site-alternating gain favored the lasing of an edge mode [4,5]. Beyond the SSH model, lasing of topological edge states protected by the inversion symmetry have also been experimentally demonstrated in a photonic nanocavity [6], as well as zero-dimensional corner states of higher-order topological models in photonic crystals [7,8,9].

Topological lasers are even more exciting in two-dimensions, where the chiral motion of the lasing mode around the sample allows to phase-lock the emission in different regions and thus maintain a high level of coherence in a spatially extended system [2]. The first realization of lasing from such chiral edge modes was in a two-dimensional magneto-optical photonic crystal under an external magnetic field [10] [Fig.1(b)], soon followed by helical edge states of a two-dimensional lattice of microring resonators [11] [Fig.1(c)]. Topological lasing, *aka* Bose-Einstein condensation, of microcavity exciton-polaritons in a two-dimensional honeycomb lattice under a strong magnetic field was reported in [12].

In spite of the impressive progress achieved in less than a decade, a number of open questions and challenges are still in front of the community: in the next paragraphs, we are going to enumerate the most exciting ones.

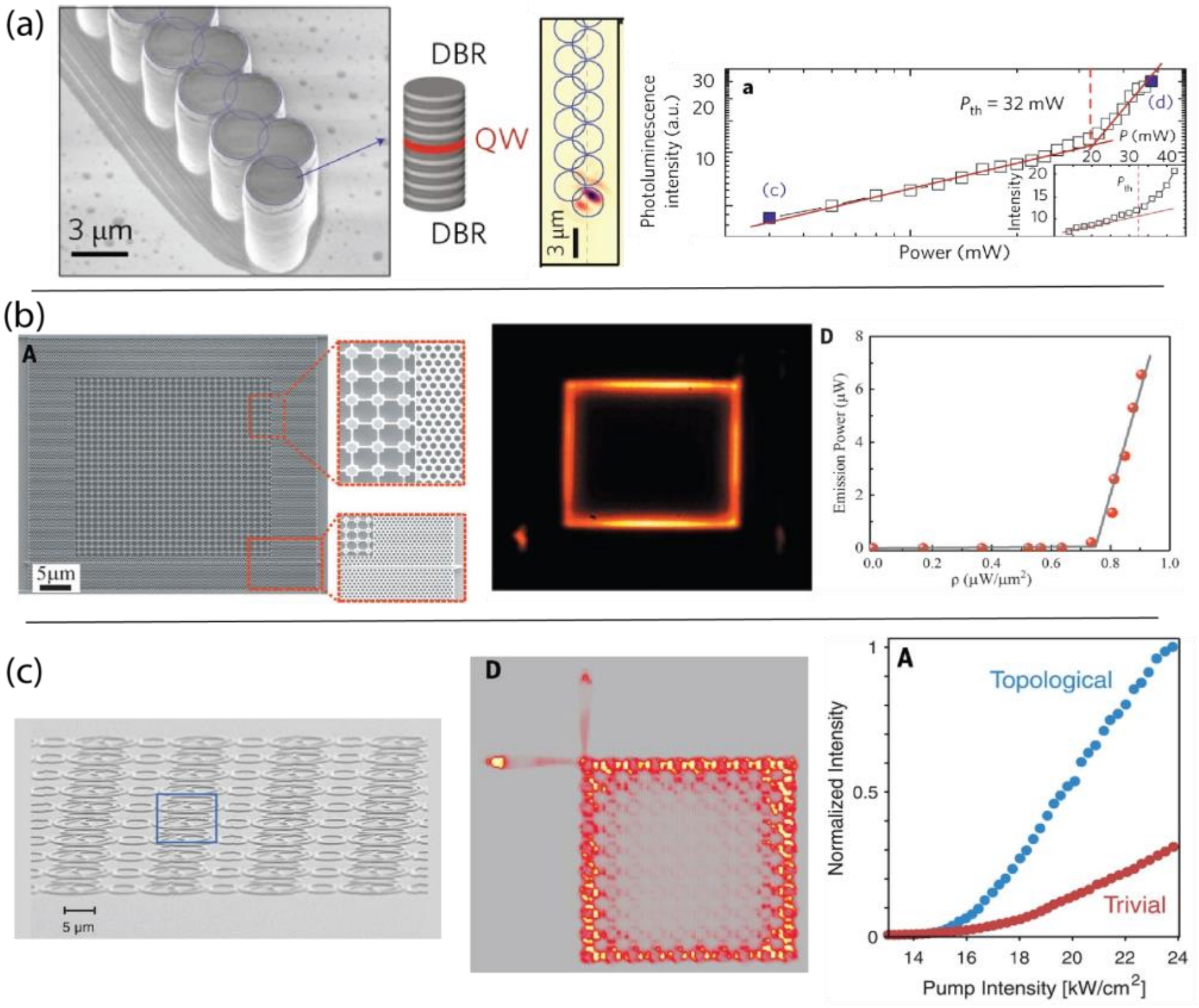

*Figure 3: Summary of some experimental realizations of topological lasers. Panel (a) is adapted from St-Jean et al. [3]. Panel (b) is adapted from Bahari et al. [10]. Panel (c) is adapted from Bandres et al. [11]. For each experiment, we show an SEM image of the device (left), the real-space profile of the laser emission (center), and the emission intensity vs. pump power curve (right).*

**Current and Future Challenges**

Although a series of experiments have confirmed topological lasing, theoretical understanding of the conditions under which a stable single mode lasing from topological edge modes can occur and of the intrinsic limitations to the coherence of such devices is far from complete, especially for the most relevant case of one-dimensional edge states of two-dimensional lattices. While the first works have already highlighted a number of novel features and unveiled some crucial advantages of topological lasing [13-16], realistic studies still need to be developed to include the peculiarities of actual devices and find strategies to tame the line-broadening and instability mechanisms due, e.g., to gain medium dynamics.

From the perspective of integrated devices for practical applications, it is typically beneficial to avoid the need for strong magnetic fields to induce a sizable topological gap. To this purpose, one needs to develop new materials with larger magneto-optical response or devise alternative strategies to ensure a robust band topology against disorder. Another important task in view of applications is to develop topological laser devices operating under an electrical rather than optical pumping, so as to avoid the need for external laser sources. A remarkable first step in this direction was the realization of a topological quantum cascade laser operating in the THz range in the edge state of a photonic valley-Hall system [17], followed by an electrically pumped spin-Hall topological laser device operating at room temperature at telecom wavelengths [18].

**Advances in Science and Technology to Meet Challenges**

In recent years, there has been a rapid development in the theoretical understanding of non-Hermitian topological band structures [19,20]. Although certain topological lasers have been understood as particular realizations of non-Hermitian topological models, developments in non-Hermitian topology and in the theory of topological lasers are still following rather independent paths. We expect that a deeper understanding of topological laser operation will require a stronger connection between these two areas. Conversely, given the central role of nonlinearities (in particular gain saturation) in any laser device, the study of topological lasing may provide unique insights into the uncharted land of non-Hermitian physics beyond the linear regime. An intriguing example of this physics is already visible in topological laser devices based on time-reversal-symmetric quantum spin-Hall configurations [11]: here, the spin symmetry associated to the direction of propagation around ring resonators is spontaneously broken in the lasing state thanks to the intrinsic nonlinearity of laser operation, which may be exploited to further protect the topological laser emission against fabrication imperfections. But even when the topological laser emission is robust, the spatially extended nature of one-dimensional topological edge states makes a complete characterization of the ultimate limits of the emission coherence a very non-trivial task. As it was realized in [15], this study goes far beyond the usual single-mode Schawlow-Townes linewidth and calls for a multi-mode theoretical description of the fluctuation dynamics using advanced concepts of non-equilibrium statistical mechanics.

From an experimental point of view, a novel strategy to reinforce the topological gap of linear topological photonic structures without the need for strong magnetic fields is to embed ferromagnetic elements into the photonic structure and make them interact with material excitations which display a sizable Zeeman effect and can be strongly coupled to light, e.g. excitons. While this might not be straightforwardly done with standard CMOS-compatible materials, intense efforts are devoted to the integration of two-dimensional materials into photonics. Beyond topological lasing, such technological advances will be of interest in other areas of topological photonics, e.g. optical isolation, as well in various other fields of contemporary research such as optical communications, optical circuitries, and even spintronics.

**Concluding Remarks**

Topological laser operation is a phenomenon at the intersection of multiple research fields such as topological band structure, non-Hermitian physics, nonlinear optics, and non-equilibrium statistical mechanics. We can therefore anticipate that further unexplored features of topological lasers will be unveiled through interdisciplinary combinations of concepts from different fields. Beyond pure science, topological lasers hold a great promise for technological applications: realizing a stable single-mode topological laser is in fact a promising strategy to exploit the chiral motion of the topological edge state and ensure a robust coherence of the emission from distant sites against fabrication imperfections and environmental noise. Integrating an electrically-pumped topological laser source into photonic circuits and high-power laser systems will therefore be of great practical value.

**Acknowledgements**

IC acknowledges financial support from the European Union FET-Open grant “MIR-BOSE” (n.737017), from the H2020-FETFLAG-2018-2020 project ”PhoQuS” (n.820392), from the Provincia Autonoma di Trento, and from the Q@TN initiative. TO acknowledges support from JSPS KAKENHI Grant No. JP20H01845, JST PRESTO Grant No. JPMJPR19L2, JST CREST Go. Number JPMJCR19T1, and RIKEN iTHEMS.

## 12 - Quantum optical effects in topological photonics

*Andrea Blanco-Redondo*

Nokia Bell Labs, USA

### Status

The enticing idea that topology can protect entanglement by protecting the entanglement-carrier particles from decoherence could have a tremendous impact on matter-based approaches to quantum computing. Photons, on the other hand, do not need to be protected from quantum decoherence, as they hardly interact with the environment, but entangled states involving multiple photons can be severely distorted due scattering and phase errors caused by fabrication imperfections. This fragility of multiphoton states is one of the problems hindering the scalability of photonic integrated approaches to the large quantum computing architectures needed to solve real problems.

In the last few years, a number of pioneering experiments have begun to tackle the question of whether topology can protect photonic quantum states in integrated platforms [1]. Some of these experiments pursue robust generation and routing of quantum states created inside the topological integrated lattice, while others explore ways of providing robust phase-matching and transport of quantum states created outside the topological chip. Following earlier single-photon experiments on bulk photonic quantum walks, topologically protected transport of quantum-dot-emitted single photons was demonstrated in a planar photonic crystal waveguide [2]. In this work, shrunken and expanded versions of a honey-comb lattice of triangular holes acquired band gaps overlapping in frequency but topologically distinct, as depicted in Fig. 1(a). Consequently, the boundary between the shrunken and expanded lattice supported two unidirectional counter-propagating modes. Chiral coupling of single photons into these modes and robust propagation through sharp bends was demonstrated.

Experiments with single photons are undoubtably interesting for applications in quantum simulation and sensing. It is, however, in the protection of fragile multiphoton states that topology could have the most impact, as these states are the basis of quantum information systems (QIS). A spectrally robust source of frequency-entangled photons was demonstrated on an aperiodic two-dimensional array of silicon ring resonators [3], see Fig. 2(b). Using a bipartite array of silicon waveguides, a recent experiment [4] has proven topological protection of features in the biphoton correlation in an $m^2$ (spatial) Hilbert space, with $m$ being the number of waveguides in the lattice. As depicted in Fig. 1(c), on-chip, nonlinearly generated biphotons were coupled to a topological boundary mode and were waveguided exhibiting robustness to coupling disorder. This idea was subsequently used to protect biphoton entanglement between two spatial modes, leveraging the protection of the propagation constant of each topological mode against coupling imperfections [5], see Fig. 1(d). Related experiments with borosilicate waveguide lattices studied the impact of topology on polarization entanglement (see [1] for a more complete literature review).

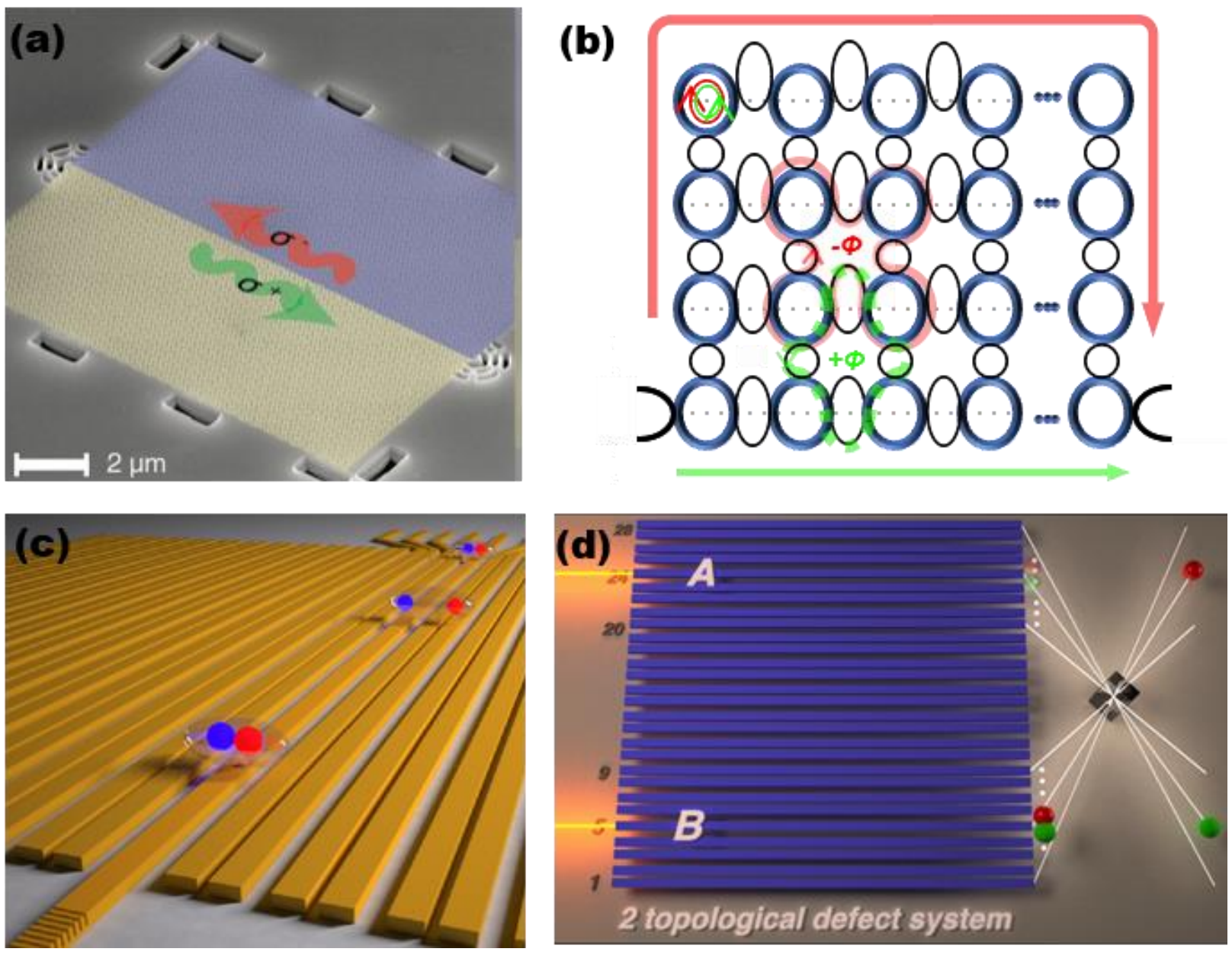

**Figure 1.** Topological nanophotonics platforms for the protection of quantum states. (a) Topologically protected chiral transport of quantum-dot-emitted single photons in a planar photonic crystal waveguide (adapted, with permission, from [2]); (b) Aperiodic bidimensional array of coupled resonators used to demonstrate the robust source of quantum light presented in [3] (figure adapted from [1]) ; (c) Bipartite array of silicon nanophotonic waveguides with one topological boundary mode for the topological protection of biphoton quantum states shown in [4]; (d) Bipartite array of silicon waveguides supporting two spatially separate topological boundary modes used to show topological protection of spatial entanglement in [5] (adapted, with permission, from [4])

## Current and Future Challenges

The study of quantum effects in topological photonics systems has just begun. The results outlined above indicate that nontrivial lattice topologies can lead to increased robustness in spectral, spatial, and polarization entanglement [1,3-5]. Nonetheless, the path for these advances in topological photonics to improve the performance of the currently developing integrated photonics quantum computing architectures is still undefined.

Some of the aforementioned topological platforms for the protection of quantum states [3-5] rely on correlated photon pair generation in silicon waveguides and rings at telecommunication wavelengths (~1550 nm) as the source of quantum light. This physical platform is compatible with the main commercial efforts in photonic quantum computing, such as those led by Psi Quantum Corp. [Fig. 2(a)] and Xanadu Quantum Technologies Inc. Further, a beamsplitter for quantum interference of topological states of light has been demonstrated [6], albeit in a borosilicate chip, providing an avenue to implement this key building block of quantum information technologies in topological systems [Fig. 2(b)]. The challenge now is to find a way to integrate lattices of many elements within the current architectures using individual waveguides, for which density of integration is a must.

Another key issue the topological photonics community must address is providing the capability to protect quantum states against the relevant types of disorder in practical QIS. Two significant obstacles to the scalability of these systems are backscattering, arising from surface roughness, and phase errors, due to variations in the widths and gaps between waveguides. In [5], Wang et al. targeted the issue of phase errors, showing coherent propagation of two entangled topological modes. Nonetheless, the robustness of this system, based on the preservation of the lattice chiral symmetry, only applied to disorder in the gaps, not in the widths. On the other hand, none of the currently available topological photonic platforms shows true protection against backscattering at optical frequencies.

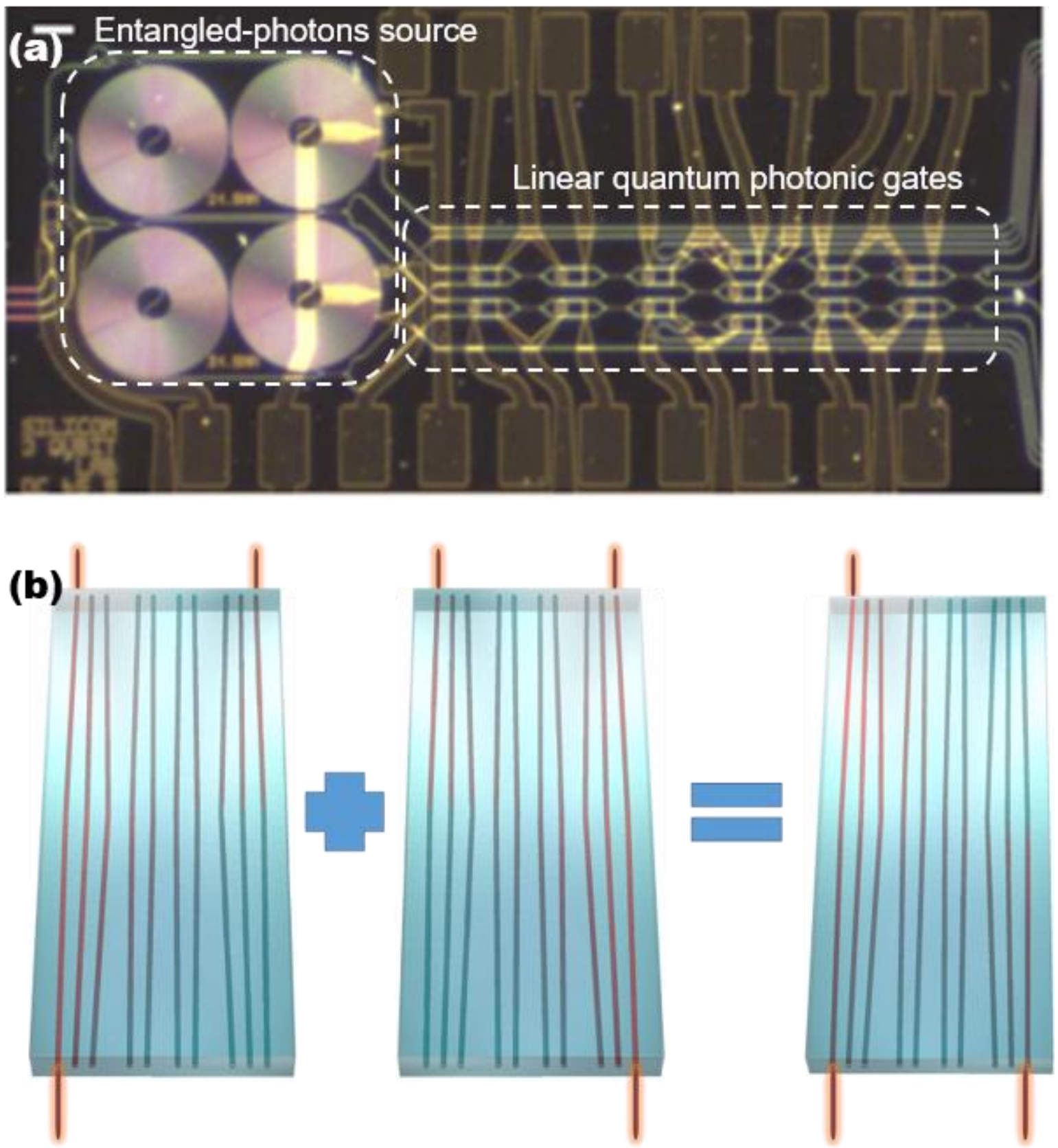

Figure 2. Key building blocks of photonic quantum information systems. (a) A silicon processor of two-qubit entangling quantum logic consisting of silicon waveguides, phase shifters, and beam splitters (adapted from Santagati *et al* 2017 *J. Opt.* **19** 114006); (b) A topological beam splitter comprising an array of coupled waveguides implementing the off-diagonal Harper model (with permission from Oded Zilberberg).

## Advances in Science and Technology to Meet Challenges

Complete immunity to disorder induced backscattering can be achieved by breaking time reversal symmetry (TRS). The weak magnetic response of materials at optical frequencies inhibits the use of magnetic fields for this purpose. It has been theoretically proposed that controlling the phase of dynamically modulated lattices of photonic resonators can lead to breaking TRS [7]. The high dynamic coupling strengths required to dominate over the decay rate of the resonator in this scheme (~ 15GHz at 1550 nm) are challenging to obtain in nanophotonics platforms, but not experimentally unattainable. The use of acoustic pumping has also been proposed as an avenue to break TRS in nanophotonic platforms with GHz bandwidths, achieving non-reciprocal propagation without magneto-optics [8]. We should note that it has also been suggested that truly backscatter-free propagation could be achieved in parity-time-duality-invariant platforms made of nonperiodic continuous media (see [1] for a description and specific works). Realistic implementations of these general ideas, adapted to be compatible with existing QIS, could lead to truly backscatter-free quantum nanophotonic platforms.

Future advances in topological orders induced by photon interactions could also have an impact in building robust QIS. These are systems where the single-particle band structure is trivial, but the two-particle bands feature topological band gaps and doublon edge states. A recent experiment in an array of superconducting qubits demonstrated the potential of interactive systems to enhance the coherent operation of superconducting qubit ensembles by reducing the parameter spread between the individually fabricated qubits [9]. The development of nonreciprocal quantum-limited travelling-wave amplifiers is also critical to superconducting qubits experiments, as these amplifiers show potential to replace the currently used circulator technology which is exceedingly lossy. A model to produce topologically protected quantum-limited amplifiers has been proposed, showing protection against internal losses and backscattering [10]. The proposed photonic device could also serve as a source of chiral squeezed states, crucial to QIS (as pursued e.g. by Xanadu Quantum Technologies Inc.) but also to quantum sensing beyond the limits usually set by quantum mechanics.

**Concluding Remarks**

The research discussed here evidences the potential of topology to protect photonic quantum states in a variety of ways, from the topological protection of multiphoton entanglement to quantum interference, all the way through the production of chiral squeezed states and robust superconducting qubits. Consequently, significant applications such as quantum computing and quantum sensing could profit from the introduction of topological concepts in their architectures. Much of this research has been done in CMOS-compatible platforms that are well-suited for scalable photonic QIS. Nonetheless, there is significant work ahead for the topological photonics and the quantum photonics technologies communities to figure out, together, if and how topological ideas can be efficiently integrated with photonic quantum information platforms and how much benefit would stem from that.

**Acknowledgements**

A.B.-R would like to acknowledge support from Nokia-Bell Labs.